\pdfoutput=1
\documentclass[12pt,a4paper]{article}

\usepackage{ifthen}
\usepackage{subcaption}
\usepackage{float}
\usepackage{multirow}
\usepackage{siunitx}
\usepackage{booktabs}
\usepackage{xstring}
\newcommand{\valinf}[1]{%
    \IfEqCase{#1}{%
        {/}{--}%
    }[\IfBeginWith{#1}{<}{\ensuremath{#1}}{\phantom{\ensuremath{<~}}#1}]%
}
\newboolean{pdflatex}
\setboolean{pdflatex}{true} 

\newboolean{articletitles}
\setboolean{articletitles}{true} 

\newboolean{uprightparticles}
\setboolean{uprightparticles}{false} 

\def \etactwos {\etac(2S)}
\def \etacones {\etac(1S)}

\def\paperauthors{LHCb collaboration} 
\def\paperasciititle{Measurement of inclusive production of charmonium states in b-hadron decays via their decay into phi phi} 
\def\papertitle{Measurement of inclusive production of charmonium states in $b$-hadron decays via their decay into $\phi \phi$} 
\def\paperkeywords{{High Energy Physics}, {LHCb}} 
\def\papercopyright{\the\year\ CERN for the benefit of the LHCb collaboration} 
\def\paperlicence{CC BY 4.0 licence}
\def\paperlicenceurl{https://creativecommons.org/licenses/by/4.0/}

\newif\ifEnableSectionTOCLinks
\EnableSectionTOCLinksfalse 

\usepackage[top=1in, bottom=1.25in, left=1in, right=1in]{geometry}

\usepackage{microtype}
\usepackage{lineno}  
\usepackage{xspace} 
\usepackage{caption} 

\usepackage{graphicx}  
\usepackage{color}
\usepackage{colortbl}
\graphicspath{{./figs/}} 

\usepackage{amsmath} 
\usepackage{amssymb}
\usepackage{amsfonts}
\usepackage{upgreek} 

\newcommand*\patchAmsMathEnvironmentForLineno[1]{%
\expandafter\let\csname old#1\expandafter\endcsname\csname #1\endcsname
\expandafter\let\csname oldend#1\expandafter\endcsname\csname
end#1\endcsname
 \renewenvironment{#1}%
   {\linenomath\csname old#1\endcsname}%
   {\csname oldend#1\endcsname\endlinenomath}%
}
\newcommand*\patchBothAmsMathEnvironmentsForLineno[1]{%
  \patchAmsMathEnvironmentForLineno{#1}%
  \patchAmsMathEnvironmentForLineno{#1*}%
}
\AtBeginDocument{%
\patchBothAmsMathEnvironmentsForLineno{equation}%
\patchBothAmsMathEnvironmentsForLineno{align}%
\patchBothAmsMathEnvironmentsForLineno{flalign}%
\patchBothAmsMathEnvironmentsForLineno{alignat}%
\patchBothAmsMathEnvironmentsForLineno{gather}%
\patchBothAmsMathEnvironmentsForLineno{multline}%
\patchBothAmsMathEnvironmentsForLineno{eqnarray}%
}

\usepackage[pdftex,
            pdfauthor={\paperauthors},
            pdftitle={\paperasciititle},
            pdfkeywords={\paperkeywords}]{hyperref}
\usepackage{hyperxmp}
\hypersetup{
    pdfcopyright={Copyright (C) \papercopyright},
    pdflicenseurl={\paperlicenceurl}
}

\usepackage[colorinlistoftodos,textsize=scriptsize]{todonotes}

\usepackage[bottom,flushmargin,hang,multiple]{footmisc}

\usepackage[all]{hypcap} 

\usepackage{xspace} 
\usepackage{upgreek}

\def\lhcb   {\mbox{LHCb}\xspace}

\def\MagUp {\mbox{\em Mag\kern -0.05em Up}\xspace}

\ifthenelse{\boolean{uprightparticles}}%
{

 \def\Peta        {\ensuremath{\upeta}\xspace}

 \def\Pmu         {\ensuremath{\upmu}\xspace}

 \def\Pchi        {\ensuremath{\upchi}\xspace}                 
 \def\Ppsi        {\ensuremath{\uppsi}\xspace}

 \def\PDelta      {\ensuremath{\Delta}\xspace}                 
 \def\PXi         {\ensuremath{\Xi}\xspace}                 
 \def\PLambda     {\ensuremath{\Lambda}\xspace}                 
 \def\PSigma      {\ensuremath{\Sigma}\xspace}                 
 \def\POmega      {\ensuremath{\Omega}\xspace}                 
 \def\PUpsilon    {\ensuremath{\Upsilon}\xspace}
 \let\oldPi\Pi
 \def\PPi         {\ensuremath{\oldPi}\xspace}

 \def\PB      {\ensuremath{\mathrm{B}}\xspace}                 
 \def\PD      {\ensuremath{\mathrm{D}}\xspace}                 
 \def\PJ      {\ensuremath{\mathrm{J}}\xspace}                 
 \def\PK      {\ensuremath{\mathrm{K}}\xspace}                 
 \def\Pb      {\ensuremath{\mathrm{b}}\xspace}                 
 \def\Pc      {\ensuremath{\mathrm{c}}\xspace}

 \def\Pp      {\ensuremath{\mathrm{p}}\xspace}                 

 \def\Ps      {\ensuremath{\mathrm{s}}\xspace}

 \def\thebaroffset{0.0em}
}
{

 \def\Peta        {\ensuremath{\eta}\xspace}

 \def\Pmu         {\ensuremath{\mu}\xspace}

 \def\Pchi        {\ensuremath{\chi}\xspace}                 
 \def\Ppsi        {\ensuremath{\psi}\xspace}                 
                  
 \mathchardef\PDelta="7101
 \mathchardef\PXi="7104
 \mathchardef\PLambda="7103
 \mathchardef\PSigma="7106
 \mathchardef\POmega="710A
 \mathchardef\PUpsilon="7107
 \mathchardef\PPi="7105
 \def\PB      {\ensuremath{B}\xspace}                 
 \def\PD      {\ensuremath{D}\xspace}                 
 \def\PJ      {\ensuremath{J}\xspace}                 
 \def\PK      {\ensuremath{K}\xspace}                 
 \def\Pb      {\ensuremath{b}\xspace}                 
 \def\Pc      {\ensuremath{c}\xspace}

 \def\Pp      {\ensuremath{p}\xspace}                 

 \def\Ps      {\ensuremath{s}\xspace}

 \def\thebaroffset{0.18em}
}
\newcommand{\offsetoverline}[2][\thebaroffset]{\kern #1\overline{\kern -#1 #2}}%

\makeatletter
\ifcase \@ptsize \relax
  \newcommand{\miniscule}{\@setfontsize\miniscule{4}{5}}
\or
  \newcommand{\miniscule}{\@setfontsize\miniscule{5}{6}}
\or
  \newcommand{\miniscule}{\@setfontsize\miniscule{5}{6}}
\fi
\makeatother

\DeclareRobustCommand{\optbar}[1]{\shortstack{{\miniscule (\rule[.5ex]{1.25em}{.18mm})}
  \\ [-.7ex] $#1$}}

\def\mup        {{\ensuremath{\Pmu^+}}\xspace}
\def\mun        {{\ensuremath{\Pmu^-}}\xspace} 

\def\mumu       {{\ensuremath{\Pmu^+\Pmu^-}}\xspace}

\def\squark    {{\ensuremath{\Ps}}\xspace}

\def\cquark    {{\ensuremath{\Pc}}\xspace}
\def\cquarkbar {{\ensuremath{\overline \cquark}}\xspace}
\def\ccbar     {{\ensuremath{\cquark\cquarkbar}}\xspace}
\def\bquark    {{\ensuremath{\Pb}}\xspace}

\def\kaon    {{\ensuremath{\PK}}\xspace}

\def\KorKbar {\kern \thebaroffset\optbar{\kern -\thebaroffset \PK}{}\xspace}

\def\Kp      {{\ensuremath{\kaon^+}}\xspace}

\def\KS      {{\ensuremath{\kaon^0_{\mathrm{S}}}}\xspace}

\def\D       {{\ensuremath{\PD}}\xspace}

\def\DorDbar {\kern \thebaroffset\optbar{\kern -\thebaroffset \PD}\xspace}

\def\Dp      {{\ensuremath{\D^+}}\xspace}
\def\Dm      {{\ensuremath{\D^-}}\xspace}

\def\DpDm    {\ensuremath{\Dp {\kern -0.16em \Dm}}\xspace}

\def\B       {{\ensuremath{\PB}}\xspace}

\def\BorBbar {\kern \thebaroffset\optbar{\kern -\thebaroffset \PB}\xspace}
\def\Bz      {{\ensuremath{\B^0}}\xspace}

\def\Bd      {{\ensuremath{\B^0}}\xspace}

\def\BdorBdbar {\kern \thebaroffset\optbar{\kern -\thebaroffset \Bd}\xspace}
\def\Bu      {{\ensuremath{\B^+}}\xspace}

\def\Bp      {{\ensuremath{\Bu}}\xspace}

\def\Bs      {{\ensuremath{\B^0_\squark}}\xspace}

\def\BsorBsbar {\kern \thebaroffset\optbar{\kern -\thebaroffset \Bs}\xspace}
\def\Bc      {{\ensuremath{\B_\cquark^+}}\xspace}

\def\jpsi     {{\ensuremath{{\PJ\mskip -3mu/\mskip -2mu\Ppsi}}}\xspace}

\def\etac     {{\ensuremath{\Peta_\cquark}}\xspace}

\def\chic     {{\ensuremath{\Pchi_\cquark}}\xspace}
\def\chiczero {{\ensuremath{\Pchi_{\cquark 0}}}\xspace}
\def\chicone  {{\ensuremath{\Pchi_{\cquark 1}}}\xspace}
\def\chictwo  {{\ensuremath{\Pchi_{\cquark 2}}}\xspace}
\def\chicJ    {{\ensuremath{\Pchi_{\cquark J}}}\xspace}

\def\Y#1S{\ensuremath{\PUpsilon{(#1S)}}\xspace}

\def\proton      {{\ensuremath{\Pp}}\xspace}
\def\antiproton  {{\ensuremath{\overline \proton}}\xspace}

\def\LorLbar     {\kern \thebaroffset\optbar{\kern -\thebaroffset \PLambda}\xspace}

\def\BF         {{\ensuremath{\mathcal{B}}}\xspace}
\def\BR         {\BF}

\newcommand{\decay}[2]{\mbox{\ensuremath{#1\!\to #2}}\xspace} 

\def\to                 {\ensuremath{\rightarrow}\xspace}

\def\eps   {{\ensuremath{\varepsilon}}\xspace}

\def\AT#1     {\ensuremath{A_{\mathrm{T}}^{#1}}\xspace}           

\def\C#1      {\ensuremath{\mathcal{C}_{#1}}\xspace}                       
\def\Cp#1     {\ensuremath{\mathcal{C}_{#1}^{'}}\xspace}                    
\def\Ceff#1   {\ensuremath{\mathcal{C}_{#1}^{\mathrm{(eff)}}}\xspace}        
\def\Cpeff#1  {\ensuremath{\mathcal{C}_{#1}^{'\mathrm{(eff)}}}\xspace}       
\def\Ope#1    {\ensuremath{\mathcal{O}_{#1}}\xspace}                       
\def\Opep#1   {\ensuremath{\mathcal{O}_{#1}^{'}}\xspace}                    

\newcommand{\nospaceunit}[1]{\ensuremath{\text{#1}}}       
\newcommand{\aunit}[1]{\ensuremath{\text{\,#1}}}       

\newcommand{\tev}{\aunit{Te\kern -0.1em V}\xspace}
\newcommand{\gev}{\aunit{Ge\kern -0.1em V}\xspace}
\newcommand{\mev}{\aunit{Me\kern -0.1em V}\xspace}
\newcommand{\kev}{\aunit{ke\kern -0.1em V}\xspace}
\newcommand{\ev}{\aunit{e\kern -0.1em V}\xspace}
 
\newcommand{\mevc}{\ensuremath{\aunit{Me\kern -0.1em V\!/}c}\xspace}
\newcommand{\gevc}{\ensuremath{\aunit{Ge\kern -0.1em V\!/}c}\xspace}
\newcommand{\mevcc}{\ensuremath{\aunit{Me\kern -0.1em V\!/}c^2}\xspace}
\newcommand{\gevcc}{\ensuremath{\aunit{Ge\kern -0.1em V\!/}c^2}\xspace}

\def\mum  {\ensuremath{\,\upmu\nospaceunit{m}}\xspace}

\def\fb   {\ensuremath{\aunit{fb}}\xspace}
\def\invfb   {\ensuremath{\fb^{-1}}\xspace}

\def\gsim{{~\raise.15em\hbox{$>$}\kern-.85em
          \lower.35em\hbox{$\sim$}~}\xspace}
\def\lsim{{~\raise.15em\hbox{$<$}\kern-.85em
          \lower.35em\hbox{$\sim$}~}\xspace}

\def\pt         {\ensuremath{p_{\mathrm{T}}}\xspace}

\def\ptot       {\ensuremath{p}\xspace}

\def\evtgen     {\mbox{\textsc{EvtGen}}\xspace}

\def\geant      {\mbox{\textsc{Geant4}}\xspace}

\def\photos     {\mbox{\textsc{Photos}}\xspace}

\def\pythia     {\mbox{\textsc{Pythia}}\xspace}

\def\tell1  {TELL1\xspace}
\def\ukl1   {UKL1\xspace}

\newcommand{\vs}{\mbox{\itshape vs.}\xspace}

\newcommand{\lhcborcid}[1]{\href{https://orcid.org/#1}{\hspace*{0.1em}\raisebox{-0.45ex}{\includegraphics[width=1em]{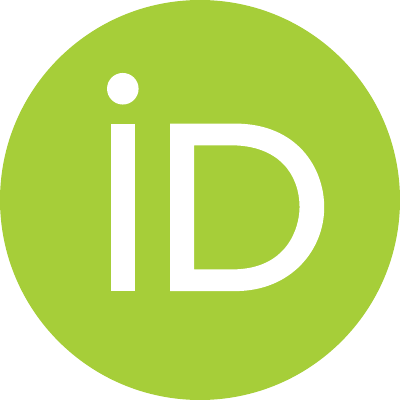}}}}

\hypersetup{
  colorlinks   = true, 
  urlcolor     = blue, 
  linkcolor    = blue, 
  citecolor    = red   
}

\ifEnableSectionTOCLinks
    \usepackage[explicit]{titlesec} 
    
    \let\oldcontentsline\contentsline
    \renewcommand

    \titleformat{\section}{\normalfont\Large\bf}{\hyperlink{tocsection.\thesection}{{\thesection} \parbox[t]{\dimexpr\textwidth-1pc}{#1}}}{1pc}{}

    \titleformat{\subsection}{\normalfont\bf}{\hyperlink{tocsubsection.\thesubsection}{{\thesubsection} \parbox[t]{\dimexpr\textwidth-1pc}{#1}}}{1pc}{}

    \titleformat{name=\section,numberless}[display]{}{}{0pt}{\normalfont\Huge\bfseries #1}
\fi

\usepackage{cite} 
\usepackage{LHCb/mciteplus}

\begin{document}

\renewcommand{\thefootnote}{\fnsymbol{footnote}}
\setcounter{footnote}{1}


\begin{titlepage}
\pagenumbering{roman}

\vspace*{-1.5cm}
\centerline{\large EUROPEAN ORGANIZATION FOR NUCLEAR RESEARCH (CERN)}
\vspace*{1.5cm}
\noindent
\begin{tabular*}{\linewidth}{lc@{\extracolsep{\fill}}r@{\extracolsep{0pt}}}
\ifthenelse{\boolean{pdflatex}}
{\vspace*{-1.5cm}\mbox{\!\!\!\includegraphics[width=.14\textwidth]{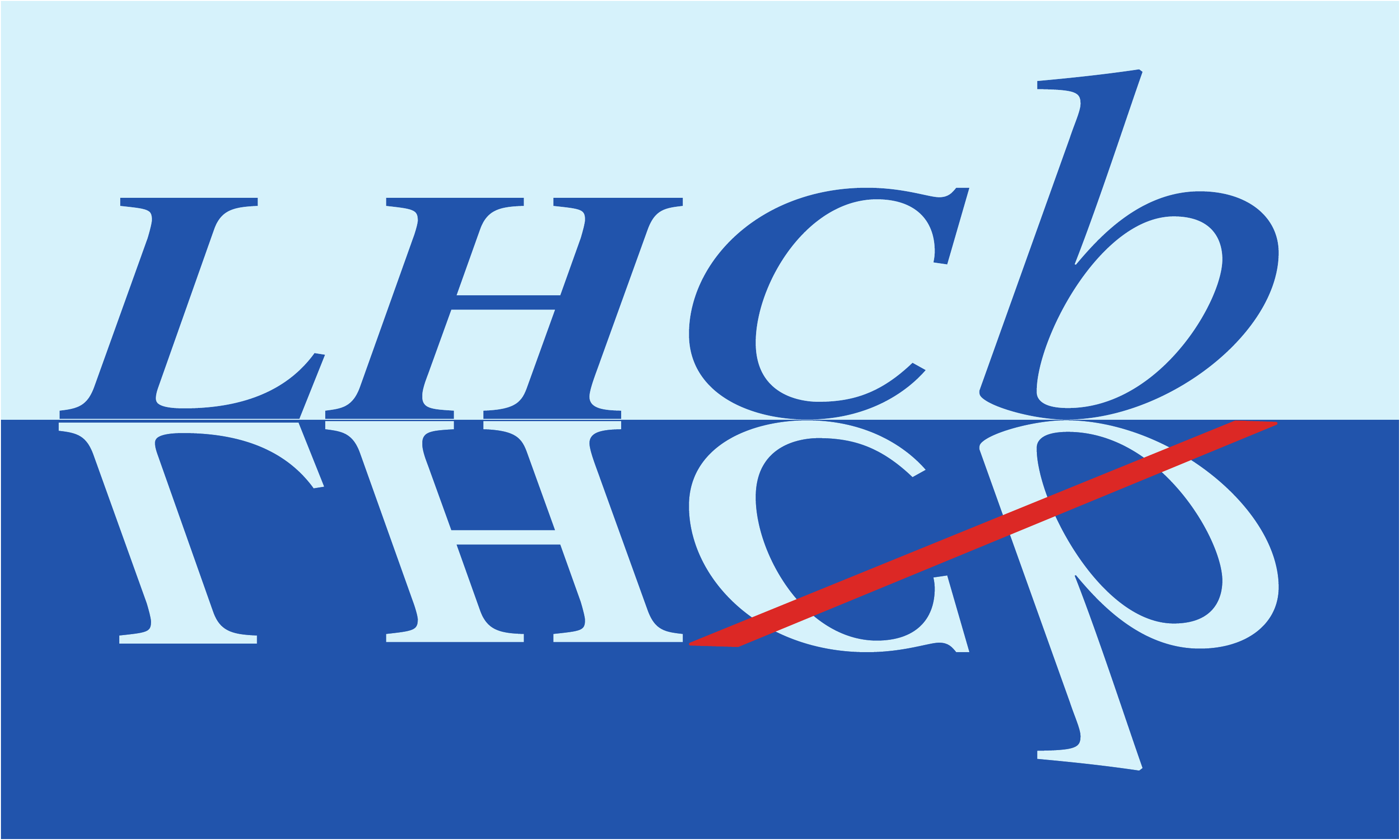}} & &}%
{\vspace*{-1.2cm}\mbox{\!\!\!\includegraphics[width=.12\textwidth]{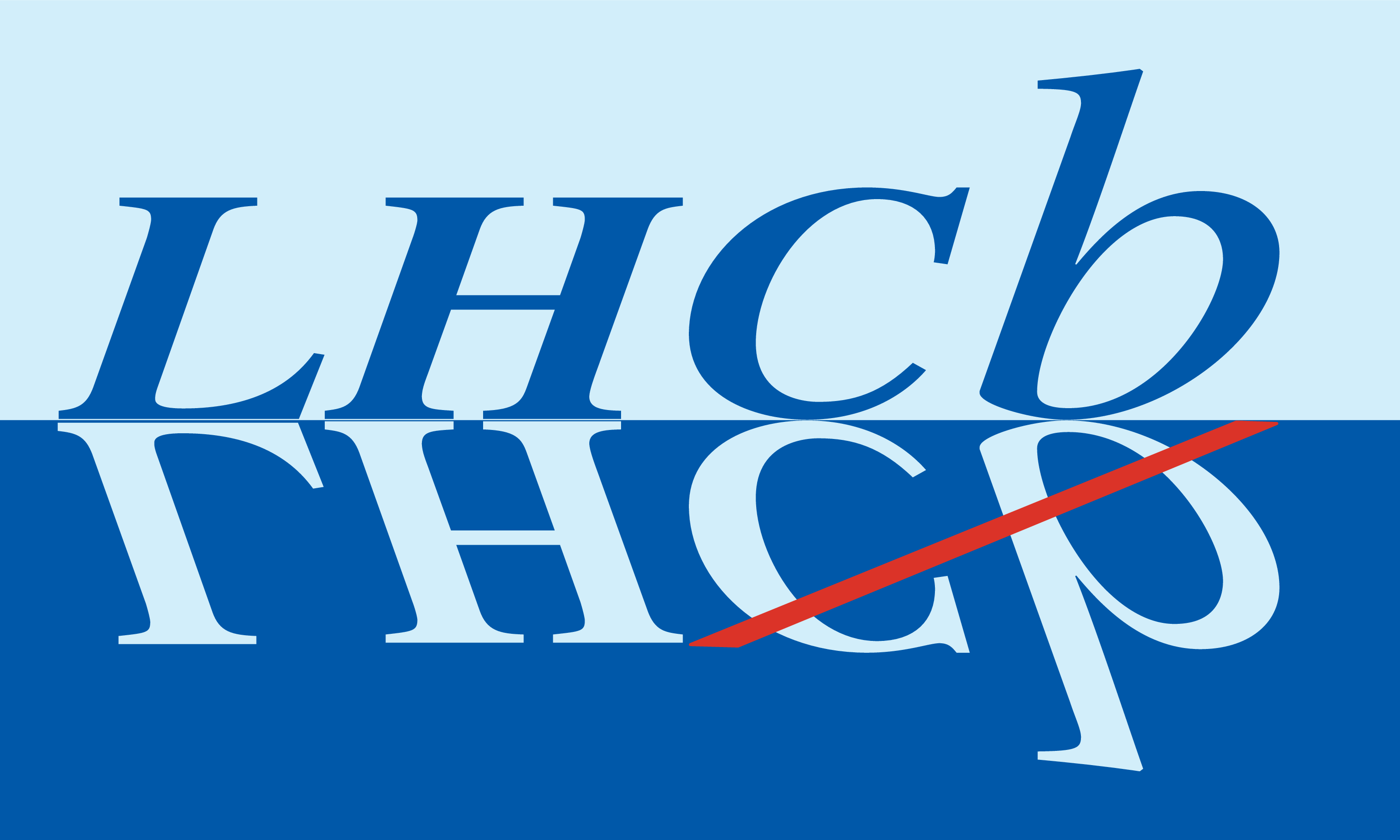}} & &}%
\\
 & & CERN-EP-2026-058 \\  
 & & LHCb-PAPER-2025-058 \\  
 & & April 9, 2026 \\ 
 & & \\
\end{tabular*}

\vspace*{3.0cm}

{\normalfont\bfseries\boldmath\huge
\begin{center}
  \papertitle 
\end{center}
}

\vspace*{1.0cm}

\begin{center}
\paperauthors\footnote{Authors are listed at the end of this paper.}
\end{center}

\vspace{\fill}

\begin{abstract}

\noindent The inclusive production of the $\etac(1S)$, $\etac(2S)$ and $\chi_{c}$ charmonium states in $b$-hadron decays is studied with LHCb Run~2 data, corresponding to an integrated luminosity of $5.9 \invfb$, using charmonia decays to $\phi\phi$ pairs. The production branching fractions of the $\chi_{c}(1P)$ states in $b$-hadron decays are measured, using \mbox{$b \to \etac(1S) (\to \phi \phi) X$} as a normalisation channel, with $X$ indicating any additional particles. The results are
\begin{align*}
    &{\cal{B}} (b \to \chi_{c0} X) = (1.34 \pm 0.13 \pm 0.06 \pm 0.37) \times 10^{-3},\\
    &{\cal{B}} (b \to \chi_{c1} X) = (1.58 \pm 0.12 \pm 0.09 \pm 0.44) \times 10^{-3},\\
    &{\cal{B}} (b \to \chi_{c2} X) = (0.55 \pm 0.08 \pm 0.05 \pm 0.15) \times 10^{-3},
\end{align*}
where the first uncertainty is statistical, the second systematic and the last is due to the limited knowledge of externally measured branching fractions. The production branching fraction of $\etac(2S)$ times the branching fraction of its decay into $\phi \phi$ is measured as \mbox{$\BR (b \to \etac(2S) X) \times \BR (\etac(2S) \to \phi \phi) = (4.0 \pm 0.6 \pm 0.6 \pm 1.1) \times 10^{-7}$}. Furthermore, the mass of the $\etac(1S)$ state is measured to be \mbox{$M_{\etac(1S)} = 2984.1 \pm 0.5 \pm 0.5\mev$} with the best precision to date.
  
\end{abstract}

\vspace*{0.5cm}

\begin{center}
  Submitted to
  Eur.~Phys.~J.~C
\end{center}

\vspace{\fill}

{\footnotesize 
\centerline{\copyright~\papercopyright. \href{\paperlicenceurl}{\paperlicence}.}}
\vspace*{2mm}

\end{titlepage}


\newpage
\setcounter{page}{2}
\mbox{~}
%
%
%
%


\renewcommand{\thefootnote}{\arabic{footnote}}
\setcounter{footnote}{0}


\cleardoublepage


\pagestyle{plain} 
\setcounter{page}{1}
\pagenumbering{arabic}


\section{Introduction}
\label{sec:intro}

The theoretical prediction of charmonium $(\ccbar)$ production branching fractions in $b$-hadron decays relies on next-to-leading-order computations of the colour singlet (CS) and colour octet~(CO) contributions in the nonrelativistic QCD (NRQCD) framework~\cite{PhysRevD.51.1125, Beneke:1998ks, Beneke_2000, Burns_2011}. The NRQCD predictions use factorisation to separate the production into a short-distance component which can be computed using perturbative QCD, and long-distance contributions, whose parameters must be constrained experimentally~\cite{PhysRevD.51.1125}. These parameters are assumed to be universal. This means that they can be computed combining cross-section results from prompt production and from $b$-hadron decays~\cite{Schuler_1999}, with spin symmetry and factorisation assumptions. For P-wave charmonia, the cross-section computations in $b$-hadron decays are not precise enough to make reliable predictions, as the CS contribution becomes very small or negative, and the CO contribution becomes dominant~\cite{Beneke:1998ks}. However, a clean test of the charmonium production models can be performed using the relative inclusive production of the $\chicJ(1P)$ states, with $J=0,1,2$ being the spin of the state, as both the CS and CO contributions are expected to be proportional to $2J+1$ at first order~\cite{Beneke:1998ks}.\footnote{In the following, $\chicJ$ refers to the $\chicJ(1P)$ mesons.}

The theoretical computations of charmonium branching fractions in $b$-hadron decays are less precise than for prompt production. Conversely, these branching fractions are easier to measure experimentally, as the background can be eliminated more easily with requirements on the decay time, due to the relatively long $b$-hadron lifetimes. Charmonium states with $J^{PC} = 1^{--}$ where $P$ is the parity and $C$ the charge-conjugation number, mostly the $\jpsi$ and $\psi(2S)$ mesons, have been studied extensively using their decays to the $\mup\mun$ final state\cite{LHCb-PAPER-2011-003,LHCB-PAPER-2015-037, LHCb-PAPER-2018-049}. States from the $\chicJ$ family can be studied via their radiative decays to $\jpsi$ mesons. Other states, such as the $\etacones$, can only be reconstructed using hadronic final states due to $C$-parity conservation. The inclusive production of $\chicJ$ and $\etactwos$ states in $b$-hadron decays has been studied by the LHCb collaboration using charmonium decays into $\phi\phi$ with Run~1 data~\cite{LHCb-PAPER-2017-007}.\footnote{In the following, $\phi$ denotes the $\phi(1020)$ meson.} The results of the Run~1 analysis were statistically limited, which motivates a new analysis based on a larger dataset.
Charmonium production in $b$-hadron decays has also been measured by recent LHCb studies using $\proton\antiproton$ final states. Measurements have been performed with 2015 and 2016 data for the $\etacones$ meson~\cite{LHCb-PAPER-2019-024}, and with 2018 data for the $\etacones$ and $\chicJ$ states~\cite{LHCb-PAPER-2024-004}.

Reconstructing charmonium states in $b$-hadron decays not only enables the measurement of their production rates, but also provides access to their fundamental properties, such as masses and widths. Precision determinations of these quantities offer stringent tests of quantum chromodynamics and improve our understanding of quarkonium spectroscopy. The most precise measurements of charmonium masses have been achieved by the LHCb and BESIII collaborations for the $\etacones$, and by the BaBar, LHCb and BESIII collaborations for $\etactwos$~\cite{LHCb-PAPER-2016-016,LHCb-PAPER-2017-007,LHCb-PAPER-2019-024,LHCb-PAPER-2022-051,BESIII:2011ab,BESIII:2012urf,BaBar:2011gos,BESIII:2023lcc}. Concerning the $\chicJ$ states, the mass measurements are dominated by the BES and E835 collaborations for $\chiczero$, and by the E760, E835 and LHCb collaborations for $\chicone$ and $\chictwo$~\cite{BES:1998ser,BES:2005bmx,FermilabE835:2002pkx,FermilabE835:2003gwi,E760:1991qim,Andreotti:2005ts,LHCb-PAPER-2017-036}. The most precise width measurements were performed by the LHCb collaboration for $\etacones$ and $\etactwos$, by the E835 collaboration for $\chiczero$, and by the E760 and E835 collaborations for $\chicone$ and $\chictwo$, as well as the LHCb collaboration for $\chictwo$~\cite{LHCb-PAPER-2022-051,FermilabE835:2002pkx,FermilabE835:2003gwi,E760:1991qim,Andreotti:2005ts,LHCb-PAPER-2017-036}.

This paper presents a study of the inclusive production of $\etacones$, $\etactwos$ and the $\chicJ$ states in $b$-hadron decays, with the charmonia decaying into a $\phi \phi$ pair.\footnote{At LHCb, all $b$-hadron species, $\Bz$, $\Bp$, $\Bs$, $\Bc$ mesons and $b$-baryons, are produced to form the analysed $b$-hadron mixture.} The $\phi$ mesons are reconstructed via their decay into the $K^+K^-$ final state. The dataset comprises $\proton\proton$ collisions recorded by the LHCb experiment in Run~2 between 2015 and 2018, corresponding to an integrated luminosity of 5.9\invfb. Measurements are performed of the branching-fraction ratios \mbox{$\BR(b \to \chic_{1,2} X)/\BR(b\to \chiczero X)$}, the  branching fractions $\BR(b \to \chicJ X)$, and the product of branching fractions $\BR(b\to \etactwos X)\times\BR(\etactwos\to\phi\phi)$, where $X$ denotes any accompanying state. Absolute production rates from $b$-hadron decays are derived using $b \to \etacones (\to \phi \phi) X$ as a normalisation channel, where the known values of $\BR(b\to \etacones X)$ and $\BR(\etacones\to \phi \phi)$ are used. The branching fraction $\BR(b\to \etactwos X)$ on its own is not measured because $\BR(\etactwos\to\phi\phi)$ is not known. As a byproduct, this paper also measures the masses of all studied states and the widths of the $\etacones$, $\chiczero$ and $\etactwos$ mesons. This paper is organised as follows: the \lhcb detector is described in Sec.~\ref{sec:detect}; the selection procedure is outlined in Sec.~\ref{sec:selection}; the fit model is explained in Sec.~\ref{sec:model}; results are given in Sec.~\ref{sec:BF_all} and~\ref{sec:MG}; and finally conclusions are drawn in Sec.~\ref{sec:conc}.

\section{LHCb detector}
\label{sec:detect}

The \lhcb detector~\cite{LHCb-DP-2008-001,LHCb-DP-2014-002} is a single-arm forward spectrometer covering the \mbox{pseudorapidity} range $2<\eta <5$, designed for the study of particles containing \bquark or \cquark quarks.
The detector used for this analysis includes a high-precision tracking system consisting of a silicon-strip vertex detector surrounding the $\proton\proton$ interaction region~\cite{LHCb-DP-2014-001}, a large-area silicon-strip detector located upstream of a dipole magnet with a bending power of about $4{\mathrm{\,T\,m}}$, and three stations of silicon-strip detectors and straw drift tubes~\cite{LHCb-DP-2017-001} placed downstream of the magnet.
The tracking system provides a measurement of the momentum, \ptot, of charged particles with a relative uncertainty that varies from 0.5\% at low momentum to 1.0\% at 200\gev.\footnote{Natural units with $\hbar=c=1$ are used throughout.} The minimum distance of a track to a primary $\proton\proton$ collision vertex (PV), the impact parameter, is measured with a resolution of $(15+29/\pt)\mum$, where \pt is the component of the momentum transverse to the beam, in\,\gev. Different types of charged hadrons are distinguished using information from two ring-imaging Cherenkov (RICH) detectors~\cite{LHCb-DP-2012-003}. Photons, electrons and hadrons are identified by a calorimeter system consisting of scintillating-pad and preshower detectors, an electromagnetic and a hadronic calorimeter. Muons are identified by a system composed of alternating layers of iron and multiwire proportional chambers~\cite{LHCb-DP-2012-002}.

The online event selection is performed by a trigger~\cite{LHCb-DP-2012-004}, which consists of a hardware stage, based on information from the calorimeter and muon systems, followed by a software stage, which applies a full event reconstruction.
The software trigger requires a reconstructed two-, three- or four-track secondary vertex. The kaon and $\phi$ candidates must have a high momentum.
The trigger either requires the tracks to be inconsistent with originating from the PV and use a multivariate algorithm~\cite{BBDT,LHCb-PROC-2015-018} to identify secondary vertices consistent with the decay of a \bquark hadron, or require the four kaon candidates to form a common vertex, with a specific range imposed on their combined mass.
Most of the used trigger lines require the mass of the kaon pairs to be consistent with the known mass of the $\phi$ meson~\cite{PDG2024}. Triggered data further undergo a centralised, offline processing step to deliver physics-analysis-ready data across the entire \lhcb physics programme~\cite{Stripping}.

Precise mass measurements require a momentum-scale calibration. The procedure is performed using samples of $\decay{\jpsi}{\mumu}$ and $\decay{\Bu}{\jpsi\Kp}$~decays collected concurrently with the~data sample used for this analysis~\cite{LHCb-PAPER-2012-048,LHCb-PAPER-2013-011}. The~relative accuracy of this procedure is estimated to be $3 \times 10^{-4}$ using samples of fully reconstructed $\Bu$, $\PUpsilon$, $\psi$ and $\KS$ mesons.

Simulation is required to model the effects of the detector acceptance and the imposed selection requirements. In the simulation, $\proton\proton$ collisions are generated using \pythia~\cite{Sjostrand:2007gs,*Sjostrand:2006za} with a specific \lhcb configuration~\cite{LHCb-PROC-2010-056}. Decays of unstable particles are described by \evtgen~\cite{Lange:2001uf}, in which final-state radiation is generated using \photos~\cite{davidson2015photos}. The interaction of the generated particles with the detector, and its response, are implemented using the \geant toolkit~\cite{Allison:2006ve, *Agostinelli:2002hh} as described in Ref.~\cite{LHCb-PROC-2011-006}.

\section{Signal selection}
\label{sec:selection}

The selection requires four kaon candidates each with $p > 5000\mev$, $\pt > 650 \mev$ and rapidity within the LHCb acceptance. The selection also uses particle identification (PID) variables based on information from the RICH subdetectors and calorimeters to ensure those candidates are identified as kaons. The $\phi$ candidates are formed combining pairs of oppositely charged kaons with mass in the range \mbox{1002--1038\mev}.
Cases in which different kaon combinations in the same event are reconstructed as $\phi$ mesons are found to be negligible.
Charmonium candidates are formed with pairs of $\phi$ candidates with mass in the range 2800--4000\mev. To select charmonia from decays of $b$ hadrons, which traverse a significant distance in the detector on average before decaying, tracks inconsistent with originating from the PV are selected. Furthermore, a minimum requirement is set on a pseudo-proper lifetime along the beam axis~\cite{LHCb-PAPER-2011-003}.

Background from duplicate tracks sharing hits in the vertex detector is rejected with a requirement on the minimum angle between each pair of tracks. Events with multiple candidates represent less than $1.5\%$ of the total, and are retained in the sample.

\begin{figure}
    \centering
    \begin{subfigure}{0.49\textwidth}
        \centering
        \includegraphics[width=\linewidth]{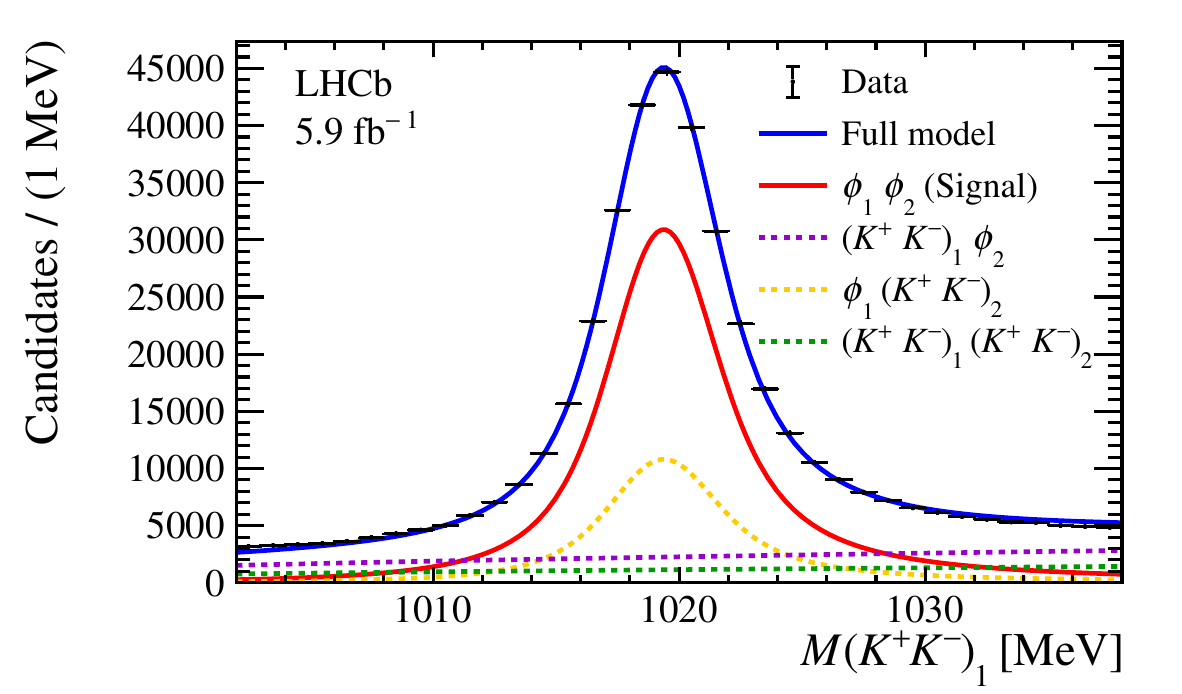}
    \end{subfigure}
    \hfill
     \begin{subfigure}{0.49\textwidth}
        \centering
        \includegraphics[width=\linewidth]{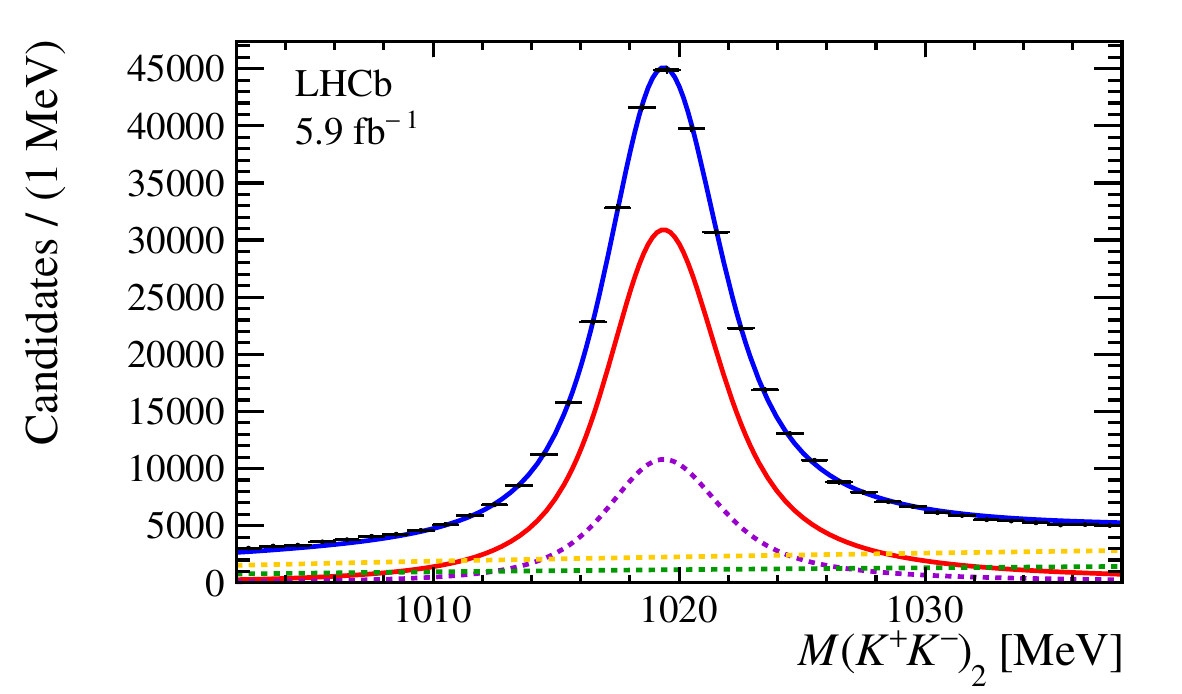}
    \end{subfigure}
    \caption{Projections of the two-dimensional fit to the two kaon-pair masses of the selected $\phi\phi$ candidates for the combined sample.}
    \label{fig:2Dfits}
\end{figure}

In order to extract the yields of $\phi \phi$ signals, two-dimensional unbinned maximum-likelihood fits to two $K^+K^-$ mass combinations are performed in 5\mev bins of the four-kaon mass combination. Each $K^+K^-$ combination is assigned randomly as the first or second $\phi$ candidate. The $\phi$ signal is modelled with a relativistic Breit--Wigner (RBW) function convolved with a Gaussian function to model detector resolution, and the background distribution is uniform. Phase-space effects are taken into account for both signal and background shapes by multiplicative factors. As both $\phi$ candidate can be background or signal, the possible combinations are the $\phi_1 \phi_2$ signal, $\phi_1 (K^+K^-)_2$, $(K^+K^-)_1 \phi_2$ and $(K^+K^-)_1 (K^+K^-)_2$. The yields of $\phi_1 (K^+K^-)_2$ and $(K^+K^-)_1 \phi_2$ backgrounds are set to be identical. The mass and width of $\phi$ mesons and width of the Gaussian detector resolution are fixed to the values obtained from a two-dimensional fit of the dikaon mass over the full $\phi \phi$ mass range, whose projections are shown in Fig.~\ref{fig:2Dfits}.
This procedure is performed for each year of the Run~2 dataset.

\section{Charmonium mass fit}
\label{sec:model}

The $\phi \phi$ mass spectra for each year, obtained from the signal yields of the two-dimensional fits, are then fitted simultaneously with a binned $\chi^2$ fit to the sum of the signal shapes of the $\etacones$, $\chiczero$, $\chicone$, $\chictwo$ and $\etactwos$ resonances, and the background. Each charmonium state is modelled by an RBW function convolved with the sum of two Gaussian functions to account for the detector resolution. The parameters of the RBW function are the mass and width of the state, the $\phi$ mass, and a radial parameter $r$, fixed to $1\,(\!\gev)^{-1}$~\cite{Blatt:1952ije}.
The ratio of the narrow and wide Gaussian widths $\sigma_\text{N}^{\ccbar}/\sigma_\text{W}^{\ccbar}$ and the fraction of narrow-to-wide Gaussian $f_{\text{NW}} = N_\text{N}/(N_\text{N} + N_\text{W})$ are shared between all states. They are determined from a fit to simulated $\chicone$ decays and fixed in the fit to data.
The ratios $\sigma_\text{N}^{\ccbar}/\sigma_\text{N}^{\etacones}$ are also determined and fixed from simulation, so that the only floated parameter for resolution functions is $\sigma_\textrm{N}^{\etacones}$.
The $\sigma_\text{N}^{\ccbar}$ values increase with the charmonium mass from around 4.5\mev for $\etacones$ to nearly 7\mev for the $\etactwos$ state.

For the $\etacones$, $\etactwos$ and $\chicone$ decays, only one orbital angular momentum value is allowed, $L=1$, $L=1$ and $L=2$, respectively. For the $\chicone$ state, this is due to the fact that $\phi\phi$ is a system of identical bosons, which prevents $L=0$ in this specific case~\cite{Yang:1950rg}. For the $\chiczero$ and $\chictwo$ states, $L=2$ is also allowed, but the lowest available value, $L=0$, is chosen. The combinatorial background is modelled with a second-order Chebyshev polynomial. The free parameters of the model are the yields and masses of all charmonium states, the natural widths of the $\etacones$, $\chiczero$ and $\etactwos$ states, the resolution $\sigma_\textrm{N}^{\etacones}$, the background yield, and the two parameters that describe its shape. The widths of the $\chicone$ and $\chictwo$ states, being much smaller than the detector resolution, are fixed to their known values~\cite{PDG2024}. The masses, widths, and resolution are shared amongst all years.

The impact of possible interference with the $\phi \phi$ nonresonant background is tested for all states. For this, a plane wave $C \exp(i\varphi)$ representing a constant background across the relevant range is allowed to interfere with the RBW function, with $C$ parametrising the amount of interfering background and $\varphi$ the interference phase. This generates three components: the RBW magnitude squared $|\mathcal{R}(x)|^2$; a pure background part $C^2$ and the interference shape $I(x)$. The pure background is absorbed in the general background description, creating the model
\begin{equation}
    \label{eq:model_wint}
    |\mathcal{R}(x)|^2 + I(x) = \frac{x \cdot \Gamma_f}{(\Delta x^2)^2 + M^2 \Gamma_f^2} + 2C \left(\frac{K(x)}{K(M)}\right)^{L} \frac{\sqrt{x \cdot \Gamma_f}\cdot (\Delta x^2\cos{\varphi} - M\Gamma_f \sin{\varphi})}{(\Delta x^2)^2 + M^2 \Gamma_f^2},
\end{equation}
where
\begin{align}
    \begin{split}
    & \Delta x^2 = x^2 - M^2, \\
    & \Gamma_f = \Gamma \cdot \left(\frac{K(x)}{K(M)}\right)^{2L+1} \cdot \left(\frac{F(rK(x))}{F(rK(M))}\right) \cdot \frac{M}{x}, \\
    & K(x) = \frac{\sqrt{\left(x^2-2m_\phi^2\right)^2-4m_\phi^4}}{2x}, \\
    & F(z) = \left\{ \begin{array}{ll}
        1, & L=0 \\
        \dfrac{1}{1 + z^2}, & L=1 \\
        \dfrac{1}{9 + 3z^2 + z^4}, & L=2
    \end{array} \right., \,\,\,\,\,\,\, z=rK(x),
    \end{split}
\end{align}
and $x$ is the $\phi\phi$ mass, $M$ the mass of the resonant state, $\Gamma$ its natural width, $m_\phi$ the known mass of the $\phi$ meson, and $r$ the radial parameter. However, absorbing $C^2$ into the general background removes a constraint on $C$, and the interference term proportional to $\sin{\varphi}$ has the same shape as the RBW function up to negligible corrections. To remove this degeneracy, the signal model is simplified by removing the $\sin{\varphi}$ term as
\begin{equation}
    \label{eq:final_model}
    \mathcal{P} \propto |\mathcal{R}(x)|^2 + I_{\text{simple}}(x) = \frac{x \cdot \Gamma_f}{(\Delta x^2)^2 + M^2 \Gamma_f^2} + 2C \left(\frac{K(x)}{K(M)}\right)^{L} \frac{\sqrt{x \cdot \Gamma_f} \cdot \Delta x^2\cos{\varphi}}{(\Delta x^2)^2 + M^2 \Gamma_f^2}.
\end{equation}
With this baseline model, the mass and width parameters can be extracted correctly, while taking interference effects into account. The drawback is that it is not possible to determine the interference phase $\varphi$, nor the contribution to the signal yield due to the interference term. To account for this, signal yields are also determined  without the interference term in the signal model of Eq.~\ref{eq:final_model} and the difference with respect to the baseline result is considered as a systematic uncertainty.

Nonresonant $\phi\phi$ background is less likely to be produced at higher $L$. To choose which states to add interference effects to, an expansion in $L$ is therefore considered, after evaluating the impact of adding interference one state at a time. The clearest sign of interference occurs for the $\etacones$ state and hints of interference for the $\etactwos$ and $\chiczero$ states are also found. This indicates that it is important to add interference for the $L=1$ states, $\etacones$ and $\etactwos$. The expansion in $L$ therefore implies that the $L=0$ states, $\chiczero$ and $\chictwo$, must also be included. The only state where no interference effects are added is $\chicone$, which can only have $L=2$. Using Wilks' theorem~\cite{Wilks:1938dza}, the interference added for the four states has a combined significance of 3.8 standard deviations ($\sigma$), considering statistical uncertainties only. 

\begin{figure}
    \centering
    \begin{subfigure}{0.49\textwidth}
        \centering
        \includegraphics[width=\linewidth]{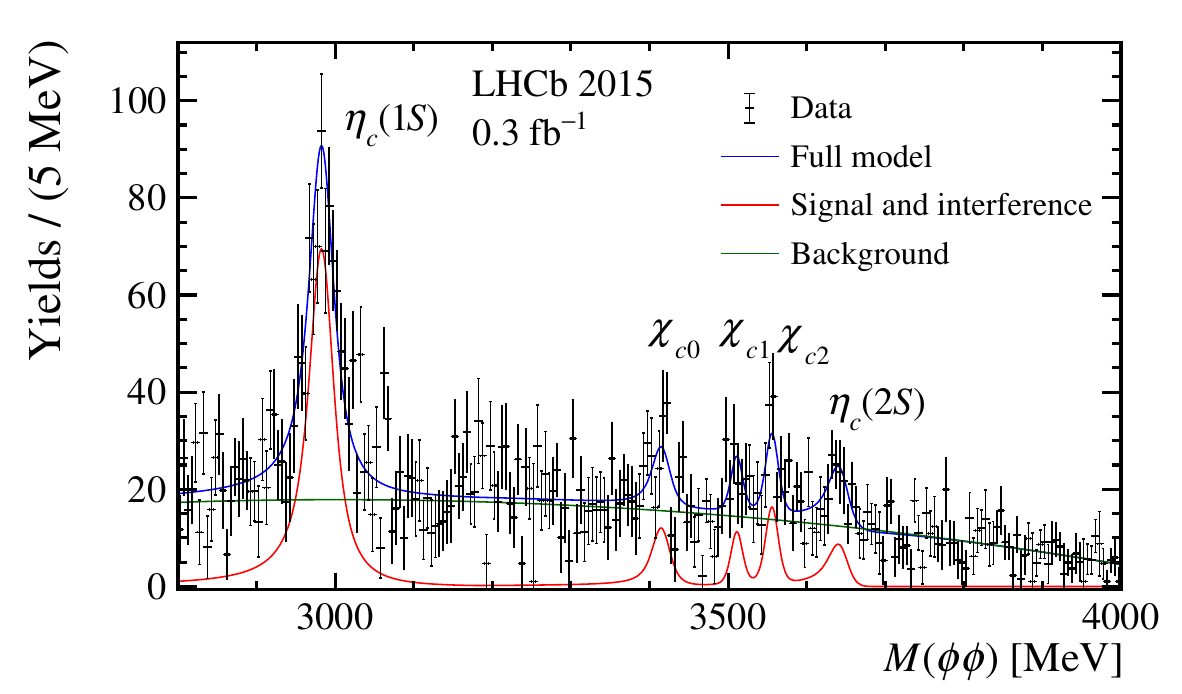}
    \end{subfigure}
    \hfill
     \begin{subfigure}{0.49\textwidth}
        \centering
        \includegraphics[width=\linewidth]{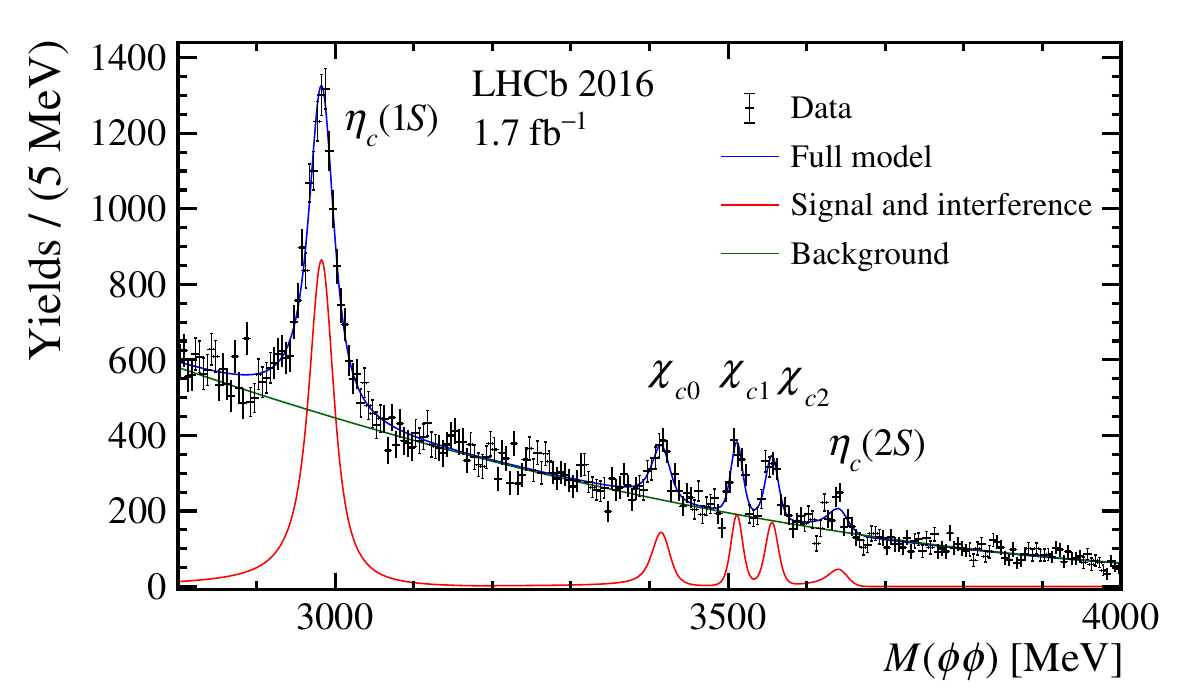}
    \end{subfigure}
    \vskip\baselineskip
    \begin{subfigure}{0.49\textwidth}
        \centering
        \includegraphics[width=\linewidth]{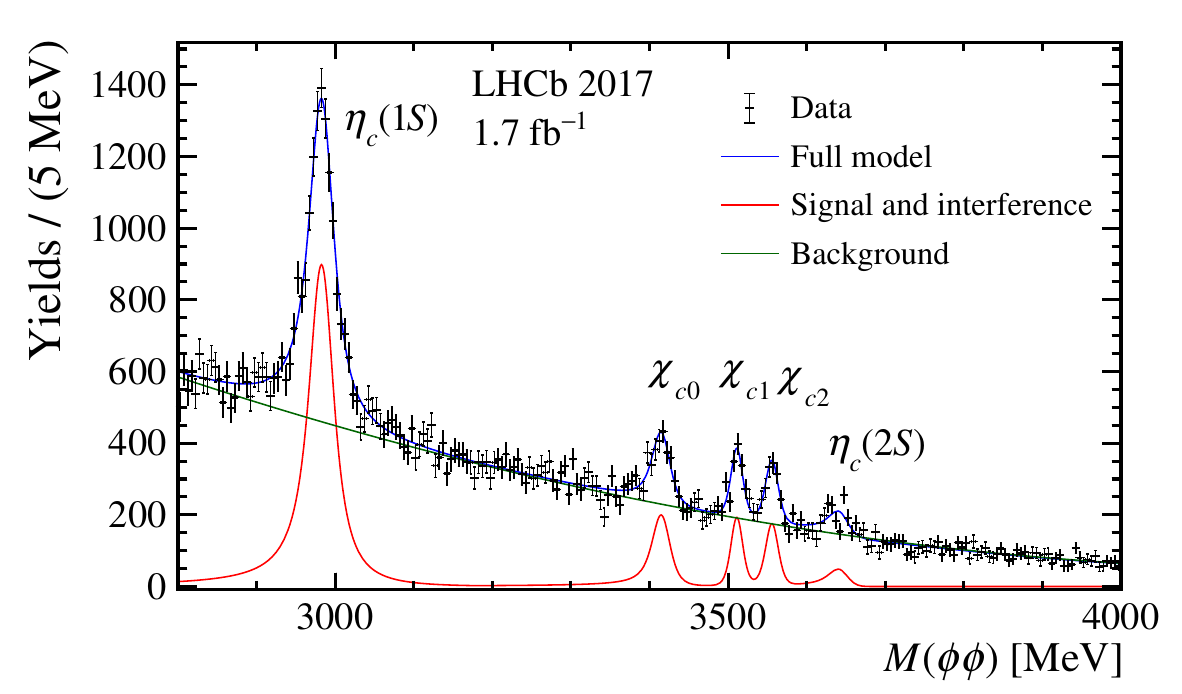}
    \end{subfigure}
    \hfill
    \begin{subfigure}{0.49\textwidth}
        \centering
        \includegraphics[width=\linewidth]{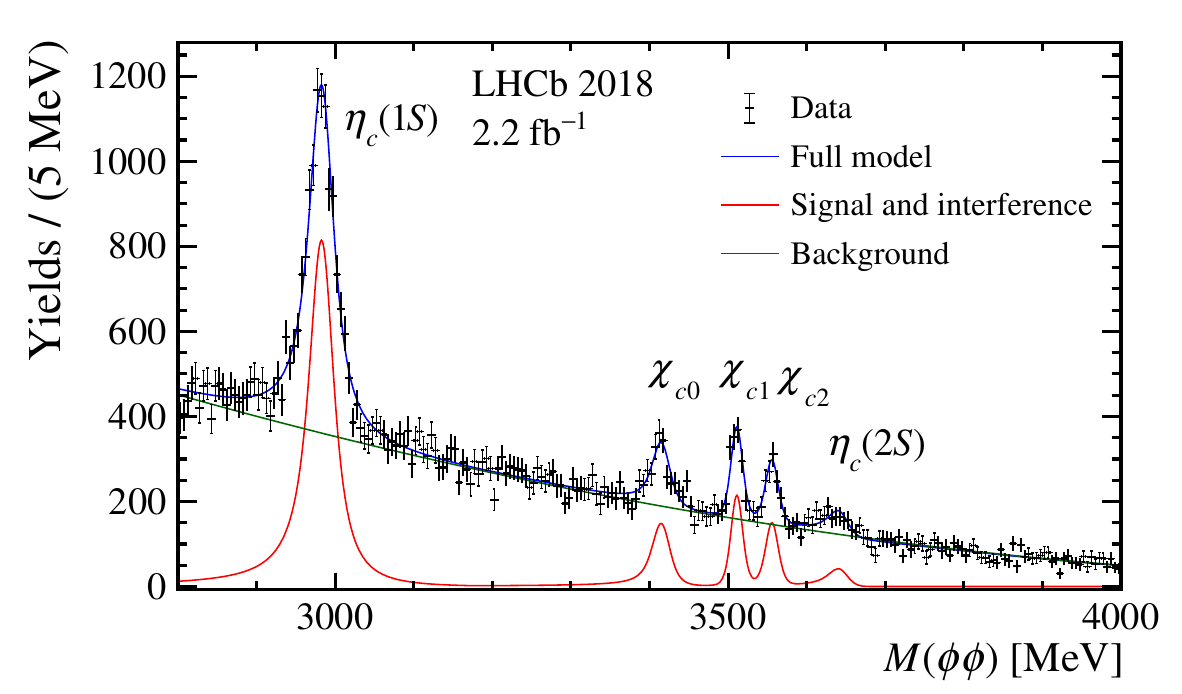}
    \end{subfigure}
    \caption{Distributions of the $\phi\phi$ mass with the result of the fit also shown.}
    \label{fig:phiphifitsint}
\end{figure}

The fits to the $\phi\phi$ mass distributions with the model described in Eq.~\ref{eq:final_model} are shown in Fig.~\ref{fig:phiphifitsint} separated by year. The extracted yield values are shown in Table~\ref{tab:yieldint}.

\begin{table}
    \centering
    \caption{Yields from the fits to the $\phi\phi$ mass distributions separated by data-taking year, where the uncertainties are statistical only.}
    \fontsize{11pt}{13.2pt}\selectfont
    \begin{tabular}{
        c
        S[table-format=3.0(2)]
        S[table-format=4.0(3)]
        S[table-format=4.0(3)]
        S[table-format=4.0(3)]
    }
        \toprule
        {State} &
        {$N^{2015}$:~$0.3\invfb$} &
        {$N^{2016}$:~$1.7\invfb$} &
        {$N^{2017}$:~$1.7\invfb$} &
        {$N^{2018}$:~$2.2\invfb$} \\
        \midrule
        $\etacones$  & 750(54)  & 9338(278) & 9700(280) & 8794(258) \\
        $\chiczero$  & 92(30)   & 1111(149) & 1550(183) & 1150(143) \\
        $\chicone$   & 42(19)   & 728(82)   & 739(83)   & 826(75)   \\
        $\chictwo$   & 70(21)   & 741(122)  & 766(133)  & 653(113)  \\
        $\etactwos$  & 90(24)   & 473(99)   & 503(108)  & 434(102)  \\
        \bottomrule
    \end{tabular}
    \label{tab:yieldint}
\end{table}

\section{Branching fractions of inclusive production in \mbox{\boldmath{$b$}-hadron} decays}
\label{sec:BF_all}

The ratios of inclusive branching fractions are extracted using the efficiency-corrected yields, $N^{\text{c}}$, as
\begin{equation}
\label{eq:BF}
    \frac{\BR (b \to A X) \times \BR (A \to \phi \phi)}{\BR (b \to B X) \times \BR (B \to \phi \phi)} = \frac{N^{\text{c}}_{A}}{N^{\text{c}}_{B}} = \frac{\sum_{\text{year}}N_A^{\text{year}}/\eps_A^{\text{year}}}{\sum_{\text{year}}N_B^{\text{year}}/\eps_B^{\text{year}}} \,,
\end{equation}
with $N_{A(B)}^{\text{year}}$ being the yield of state $A$ ($B$) for this specific year extracted from the fit, and $\eps_{A(B)}^{\text{year}}$ the corresponding efficiency determined from simulation.

\subsection{Efficiencies}
\label{sec:eff}

Efficiencies are determined from simulation and include the effects of acceptance, reconstruction, trigger, selection and the two-dimensional fits. To account for differences between data and simulation in tracking and PID efficiencies, corrections are applied based on calibration samples~\cite{LHCb-DP-2013-002,LHCb-PUB-2016-021}. Potential differences in the simulation of the global event description and the kinematic distributions could influence efficiencies and their ratios. The trigger efficiencies are lower in 2018 than for the other years, due to the introduction of more stringent requirements. Charmonium states with lower mass, in particular the $\etacones$ state, release less energy, which leads to lower efficiency as more events are lost at low kaon $\pt$. This effect is more prominent for 2015 due to trigger limitations. The efficiencies are summarised in Table~\ref{tab:eff_tot}, where the uncertainties are due to the simulation sample sizes.

\begin{table}[t]
    \centering
    \caption{\centering Total efficiencies per data-taking year for each state.}
\begin{tabular}{ccccc}
\toprule
 State
 & $\varepsilon^{2015}_{\text{tot}}~[\%]$
 & $\varepsilon^{2016}_{\text{tot}}~[\%]$
 & $\varepsilon^{2017}_{\text{tot}}~[\%]$
 & $\varepsilon^{2018}_{\text{tot}}~[\%]$ \\
\midrule
 $\etacones$     
 & $0.161 \pm 0.004$ 
 & $0.388 \pm 0.007$ 
 & $0.441 \pm 0.008$ 
 & $0.275 \pm 0.006$ \\
 $\chiczero$ 
 & $0.249 \pm 0.005$ 
 & $0.456 \pm 0.009$ 
 & $0.505 \pm 0.009$ 
 & $0.343 \pm 0.007$ \\
 $\chicone$  
 & $0.279 \pm 0.006$ 
 & $0.465 \pm 0.009$ 
 & $0.534 \pm 0.010$ 
 & $0.339 \pm 0.007$ \\
 $\chictwo$  
 & $0.272 \pm 0.006$ 
 & $0.475 \pm 0.009$ 
 & $0.525 \pm 0.010$ 
 & $0.336 \pm 0.007$ \\
 $\etactwos$ 
 & $0.317 \pm 0.006$ 
 & $0.516 \pm 0.009$ 
 & $0.556 \pm 0.010$ 
 & $0.384 \pm 0.007$ \\
 \bottomrule
\end{tabular}
\label{tab:eff_tot}
\end{table}

\subsection{Systematic uncertainties}
\label{sec:syst_BF}

Several sources of systematic uncertainties on the charmonium yield ratios are considered: the detector resolution model and constraints; potential contributions from other resonances; the choice of background model in the $\phi \phi$ mass fit; the efficiency-corrected signal yield ratios; the choice of free mass shape parameters for the $\chicone$ and $\chictwo$ contributions in the fit; the choice of orbital angular momentum for the $\chiczero$ and $\chictwo$ states; the choice of radial parameter in the RBW parametrisation; the choice of bin width; and the simplification of the interference model. For each source, 500 pseudoexperiments are generated using the baseline model, and fitted using both the baseline and alternative models. The effects of these are discussed in the following.

The impact of the resolution model is evaluated by parametrising the detector resolution with a single Gaussian function in the $\phi\phi$ mass fit, with the ratios of the Gaussian widths between the different states fixed to those obtained in the baseline fit. The robustness of this assumption is assessed by varying these ratios. To this end, the resolution is studied as a function of the energy released for each state and parametrised with a linear function, whose fit results are then used to determine alternative resolution ratios between states. Additional contributions from the $\chicone(3872)$ and $\chiczero(3915)$ resonances are included to evaluate the corresponding systematic uncertainty.
The impact of the background parametrisation in the $\phi \phi$ mass fit is evaluated by adding a third-order term to the Chebyshev polynomial. The systematic uncertainty on efficiencies is obtained by computing those from the entire simulated sample after selection, without performing the two-dimensional fits in $\phi \phi$ mass. The masses of $\chicone$ and $\chictwo$ mesons are known with more than fifteen times better precision than what can be obtained in this analysis~\cite{PDG2024}. For this reason, the systematic uncertainty due to the description of the $\chicone$ and $\chictwo$ peaks is obtained by fixing their masses to their known values.

In the baseline fit, the lowest angular momentum of the $\phi\phi$ state in each charmonium decay is taken for the RBW model when there are two options as explained in Sec.~\ref{sec:model}. The effect of this choice is evaluated by taking $L=2$ instead of $L=0$ in the $\chiczero$ and $\chictwo$ models. The uncertainty associated with the shape of the RBW is determined by varying the radial parameter $r$. It is fixed to $0.5\,(\!\gev)^{-1}$, then $3\,(\!\gev)^{-1}$~\cite{Blatt:1952ije}, and the largest difference between the two for each relevant observable is taken as the associated systematic uncertainty. The uncertainty associated with the choice of bin width in $\phi\phi$ mass is evaluated by changing the bin width from $5$ to $10 \mev$. The impact of choosing a baseline model with interference is evaluated by performing the fit with no interference in the model, as explained in Sec.~\ref{sec:model}.

Another source of uncertainty arises from potential remaining prompt charmonium contributions to the selected $\phi\phi$ candidates.
Using simulation, the selection efficiency ratio of prompt production to that of $b$-hadron decays is determined for all years to be at the level of $0.1\%$.
The ratio of prompt \vs nonprompt production is then determined to evaluate the remaining prompt contribution. No study has been performed previously in the same $\pt$ range as this analysis, nor for the $\chiczero$ or $\etactwos$ mesons. However, across all available studies on charmonium states, the ratio of prompt \vs nonprompt production is roughly 10 or lower~\cite{LHCb-PAPER-2011-003,LHCB-PAPER-2015-037,LHCb-PAPER-2018-049,LHCb-PAPER-2019-024,LHCb-PAPER-2024-004,ATLAS:2014ala}. Making the same assumption here, a $2\%$ uncertainty is added on all ratios to account for possible prompt contamination, which has a small effect on the total systematic uncertainties.

A set of 500 pseudoexperiments is used to validate the fit procedure. As a result, corrections are applied amounting to approximately 1--5$\%$ on the yield ratios, less than 0.01$\%$ on the masses and 0.1--1$\%$ on the widths. A cross-check is performed, changing the constant part of the two-dimensional fit background to an exponential function. The fit is performed in each bin, the distribution is extracted and the yield ratios are measured as explained in Sec.~\ref{sec:model}. As the difference with using the constant background is negligible, no systematic uncertainty is assigned. A cross-check is performed by repeating the analysis in bins of $\pt$, $\eta$, track multiplicity and $\phi$ angular distributions and the results do not show any significant deviation.

The systematic uncertainties from all sources for the efficiency-corrected yield ratios are given in Table~\ref{tab:syst} for the combined sample. The efficiencies presented in Sec.~\ref{sec:eff}, are included in the quoted uncertainties. The total systematic uncertainties are obtained by summing in quadrature the individual components. The dominant source of systematic uncertainty is associated with the removal of interference effects. Changing the orbital angular momentum associated with the $\chiczero$ and $\chictwo$ states is another large source.

\begin{table}
    \centering
    \caption{Relative systematic uncertainties on the efficiency-corrected signal yield ratios for each source, in percent.}
    \begin{tabular}{lcccccc}
        \toprule
        \multirow{2}{*}{Source} & \multirow{2}{*}{$\frac{\chiczero}{\etacones}$} & \multirow{2}{*}{$\frac{\chicone}{\etacones}$} & \multirow{2}{*}{$\frac{\chicone}{\chiczero}$} & \multirow{2}{*}{$\frac{\chictwo}{\etacones}$} & \multirow{2}{*}{$\frac{\chictwo}{\chiczero}$} & \multirow{2}{*}{$\frac{\etactwos}{\etacones}$} \\
        & & & & & & \\
        \midrule
        Single Gaussian resolution & \valinf{0.9} & \valinf{2.5} & \valinf{1.6} & \valinf{2.8} & \valinf{3.6} & \valinf{0.4} \\
        $\sigma_{\ccbar}/\sigma_{\etacones}$ from fit & \valinf{<0.1} & \valinf{0.7} & \valinf{0.6} & \valinf{0.5} & \valinf{0.5} & \valinf{0.3} \\
        Including $\chicone(3872)$ and $\chiczero(3915)$ & \valinf{<0.1} & \valinf{0.2} & \valinf{0.2} & \valinf{0.5} & \valinf{0.5} & \valinf{1.0} \\
        New $\phi \phi$ mass fit background & \valinf{<0.1} & \valinf{<0.1} & \valinf{0.2} & \valinf{0.5} & \valinf{0.6} & \valinf{0.1} \\
        Alternative efficiencies & \valinf{0.7} & \valinf{1.3} & \valinf{0.7} & \valinf{2.2} & \valinf{1.6} & \valinf{1.4} \\
        $M_\chicone$, $M_\chictwo$ at known values & \valinf{0.2} & \valinf{1.7} & \valinf{1.4} & \valinf{4.6} & \valinf{4.9} & \valinf{0.4} \\
        Changing $L$ for $\chiczero$ and $\chictwo$ & \valinf{3.8} & \valinf{0.2} & \valinf{3.7} & \valinf{1.6} & \valinf{2.1} & \valinf{0.7} \\
        New RBW radial parameter & \valinf{0.2} & \valinf{0.3} & \valinf{0.2} & \valinf{0.3} & \valinf{0.1} & \valinf{1.7} \\
        $10 \mev$ bin width & \valinf{0.2} & \valinf{0.1} & \valinf{0.3} & \valinf{0.3} & \valinf{0.2} & \valinf{0.3} \\
        No interference & \valinf{1.6} & \valinf{4.4} & \valinf{5.0} & \valinf{5.1} & \valinf{5.1} & \!\!\!\valinf{15.8} \\
        Residual prompt contribution & \valinf{2.0} & \valinf{2.0} & \valinf{2.0} & \valinf{2.0} & \valinf{2.0} & \valinf{2.0} \\
        \midrule
        Total systematic uncertainties& \valinf{4.7} & \valinf{5.9} & \valinf{7.0} & \valinf{8.2} & \valinf{8.6} & \!\!\!\valinf{16.2} \\
        \midrule
        Statistical uncertainties & \valinf{9.7} & \valinf{7.8} & \!\!\!\valinf{11.3} & \!\!\!\valinf{13.8} & \!\!\!\valinf{16.8} & \!\!\!\valinf{15.4} \\
        \bottomrule
    \end{tabular}
    \label{tab:syst}
\end{table}

\subsection{Results}
\label{sec:BF}

The efficiency-corrected yield ratios, which are consistent across the datasets, are shown in Table~\ref{tab:effyieldrat}. The relative branching fractions of the inclusive production of $\chicJ$ states in $b$-hadron decays are then obtained using the world averages \mbox{$\BR (\chiczero \to \phi \phi) = (8.48 \pm 0.31) \times 10^{-4}$}, \mbox{$\BR (\chicone \to \phi \phi) = (4.26 \pm 0.21) \times 10^{-4}$}, \mbox{$\BR (\chictwo \to \phi \phi) = (1.23 \pm 0.07) \times 10^{-3}$}~\cite{PDG2024}. 

\begin{table}
    \centering
    \caption{Efficiency-corrected yield ratios according to Eq.~\ref{eq:BF}, where the first uncertainty is statistical and the second systematic.}
    \begin{tabular}{cc}
        \toprule
        Ratio & Value \\
        \midrule
        $N^{\text{c}}_{\chicone}/N^{\text{c}}_{\chiczero}$ & $0.591 \pm 0.067 \pm 0.041$ \\
        $N^{\text{c}}_{\chictwo}/N^{\text{c}}_{\chiczero}$ & $0.592 \pm 0.100 \pm 0.051$ \\
        $N^{\text{c}}_{\chiczero}/N^{\text{c}}_{\etacones}$ & $0.113 \pm 0.011 \pm 0.005$ \\
        $N^{\text{c}}_{\chicone}/N^{\text{c}}_{\etacones}$ & $0.067 \pm 0.005 \pm 0.004$ \\
        $N^{\text{c}}_{\chictwo}/N^{\text{c}}_{\etacones}$ & $0.067 \pm 0.009 \pm 0.005$ \\
        $N^{\text{c}}_{\etactwos}/N^{\text{c}}_{\etacones}$ & $0.039 \pm 0.006 \pm 0.006$ \\
        \bottomrule
    \end{tabular}
    \label{tab:effyieldrat}
\end{table}

Absolute branching fractions of $\chic$ production in $b$-hadron decays are also obtained, using yield ratios with respect to the $\etacones$ state. The derivation uses the known values of \mbox{$\BR (b \to \etacones X) = (5.6 \pm 0.9) \times 10^{-3}$} and \mbox{$\BR (\etacones\to \phi \phi) = (1.8 \pm 0.4) \times 10^{-3}$}~\cite{PDG2024}. For the $\etactwos$ branching fractions, there is no measurement of \mbox{$\BR(\etactwos \to \phi \phi)$}, so only the product of \mbox{$\BR(\etactwos \to \phi \phi)$} and \mbox{$\BR (b \to \etactwos X)$} can be determined.

\begin{figure}
    \centering
    \includegraphics[width=1\linewidth]{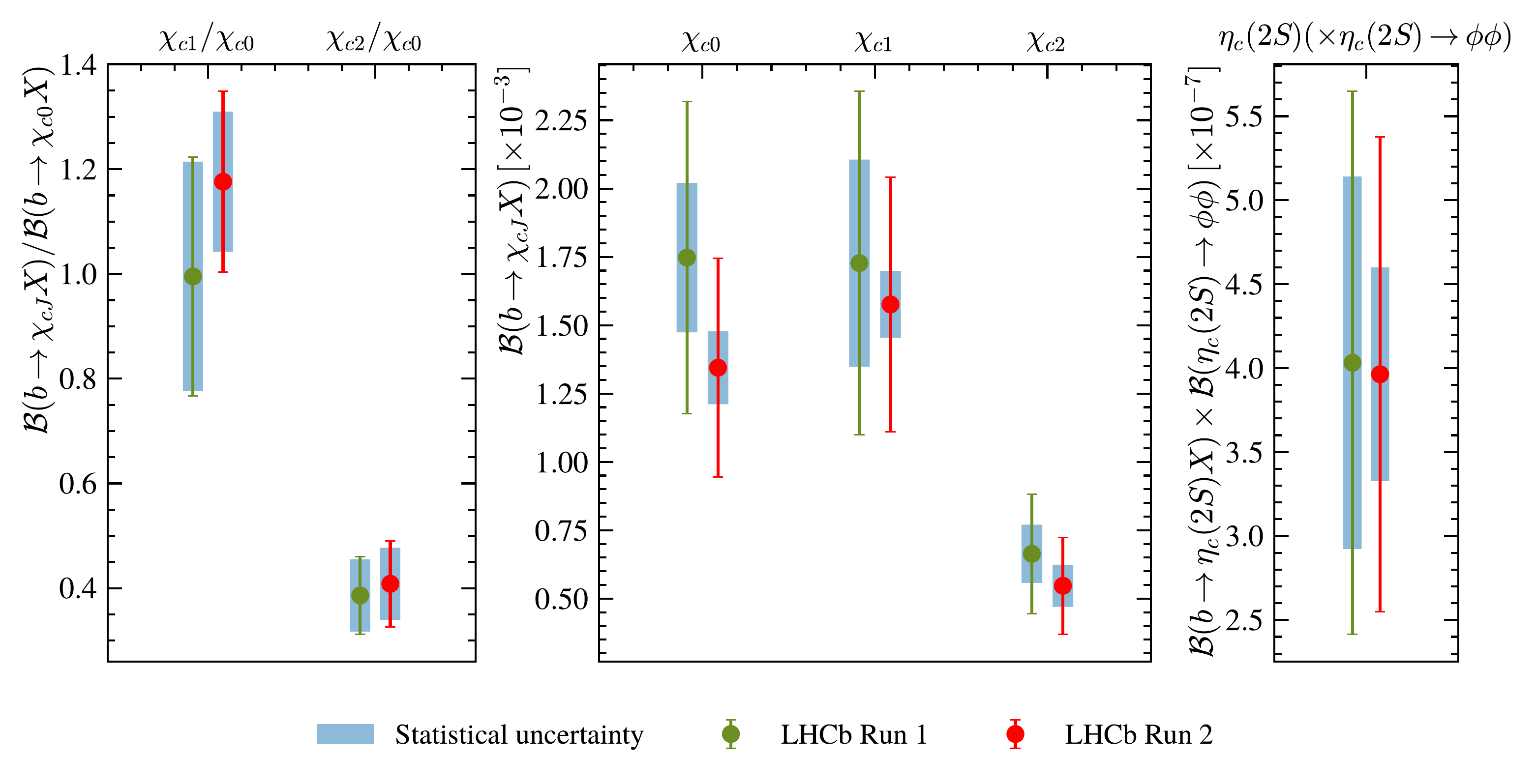}
    \caption{Measured values of (left) branching-fraction ratios of $\chi_{cJ}$ states, (middle) absolute branching fractions of $\chi_{cJ}$ states and (right) branching-fraction product with ${\cal{B}}(\etac(2S) \to \phi\phi)$ for the $\etac(2S)$ state, in orange. The corresponding Run~1 results~\cite{LHCb-PAPER-2017-007} are given in blue.}
    \label{fig:BR}
\end{figure}

The resulting branching fractions are given in Table~\ref{tab:BR}, where the third uncertainty is due to the finite precision of the known branching fractions used in the computation. The branching fractions are also shown in Fig.~\ref{fig:BR}. The uncertainties on the measured relative branching fractions between the $\chicJ$ states are dominated by the statistical uncertainties, thereby motivating improvement using larger datasets. For the absolute branching fractions, the dominant uncertainty is that associated with the knowledge of the branching fraction of the $\etacones$ decay. The Run~2 results are consistent with those from the Run~1 analysis~\cite{LHCb-PAPER-2017-007}, which are also given in Table~\ref{tab:BR} and Fig.~\ref{fig:BR}. The larger systematic uncertainties in Run~2 are due to considering a more complete model including interference effects.

This is the most precise measurement of $b$-hadron decays into the $\chictwo$ state from this admixture of $b$-hadrons. The measurement of $b$-hadron decays into the $\chicone$ state has similar precision to the most precise one, and that into the $\chiczero$ state is slightly less precise~\cite{LHCb-PAPER-2024-004}. The limitations on precision are due primarily to the uncertainty on $\BR (\etacones\to \phi \phi)$.

\begin{table}
\centering
\caption{Branching fractions measured in this analysis along with Run~1 results~\cite{LHCb-PAPER-2017-007}, where the first uncertainty is statistical, the second systematic and the third is due to the finite precision of the branching fractions used in the normalisation~\cite{PDG2024}.}
\begin{tabular}{ccc}
    \toprule
    & Run~1 & Run~2 \\
    \midrule
    $\mathcal{B} (b \to \chicone X)/\mathcal{B} (b \to \chiczero X)$ & $1.00 \pm 0.22 \pm 0.02 \pm 0.06$ & $1.18 \pm 0.13 \pm 0.08 \pm 0.07$ \\
    $\mathcal{B} (b \to \chictwo X)/\mathcal{B} (b \to \chiczero X)$ & $0.39 \pm 0.07 \pm 0.01 \pm 0.03$ & $0.41 \pm 0.07 \pm 0.04 \pm 0.03$ \\
    \midrule
    $\mathcal{B} (b \to \chiczero X) [\times 10^{-3}]$ & $1.75 \pm 0.27 \pm 0.13 \pm 0.48$ & $1.34 \pm 0.13 \pm 0.06 \pm 0.37$ \\
    $\mathcal{B} (b \to \chicone X) [\times 10^{-3}]$ & $1.73 \pm 0.38 \pm 0.14 \pm 0.48$ & $1.58 \pm 0.12 \pm 0.09 \pm 0.44$ \\
    $\mathcal{B} (b \to \chictwo X) [\times 10^{-3}]$ & $0.66 \pm 0.11 \pm 0.04 \pm 0.19$ & $0.55 \pm 0.08 \pm 0.05 \pm 0.15$ \\
    \midrule
    $\mathcal{B} (b \to \etactwos X) \times$ & \multirow{2}{*}{$4.0~\, \pm 1.1~\, \pm 0.4~\, \pm 1.1~\,$} & \multirow{2}{*}{$4.0~\, \pm 0.6~\, \pm 0.6~\, \pm 1.1~\,$} \\
    $\mathcal{B} (\etactwos \to \phi \phi) [\times 10^{-7}]$ & & \\
    \bottomrule
\end{tabular}
\label{tab:BR}
\end{table}

\section{Masses and widths}

\label{sec:MG}

The same fit to the $\phi\phi$ mass is also used to extract the masses and natural widths of the investigated states. Most sources of systematic uncertainties are the same as those discussed in Sec.~\ref{sec:syst_BF}. An additional contribution due to the momentum scaling is considered.
To evaluate this uncertainty, two small scale factors are applied to the kaon momenta in simulation, $\pm 3\times 10^{-4}$~\cite{LHCb-DP-2023-003}, and the masses and widths are extracted from a fit in both cases.

For each state, the maximum absolute difference with the result obtained from simulation without scaling is taken as a systematic uncertainty.

The values of all systematic uncertainties are given in Table~\ref{tab:syst_M} for the masses and Table~\ref{tab:syst_G} for the widths, where the totals are obtained by summing the individual components in quadrature. The largest source of uncertainty for the masses is from the momentum scaling. For the widths, the change in resolution model to a single Gaussian function constitutes the largest source.

\begin{table}[b]
    \centering
    \caption{Relative systematic uncertainties on the charmonium-state masses, in percent.}
    \begin{tabular}{lccccc}
        \toprule
        Source & $M_{\etacones}$ & $M_{\chiczero}$ & $M_{\chicone}$ & $M_{\chictwo}$ & $M_{\etactwos}$ \\
        \midrule
        Single Gaussian resolution & \valinf{<0.001} & \valinf{<0.001} & \valinf{<0.001} & \valinf{0.009} & \valinf{0.005} \\
        $\sigma_{\ccbar}/\sigma_{\etacones}$ from fit & \valinf{<0.001} & \valinf{<0.001} & \valinf{<0.001} & \valinf{<0.001} & \valinf{<0.001} \\
        With $\chicone(3872)$ and $\chiczero(3915)$ & \valinf{<0.001} & \valinf{<0.001} & \valinf{<0.001} & \valinf{<0.001} & \valinf{0.005} \\
        New $\phi \phi$ mass fit background & \valinf{<0.001} & \valinf{<0.001} & \valinf{<0.001} & \valinf{<0.001} & \valinf{0.002} \\
        $M_\chicone$, $M_\chictwo$ at known values & \valinf{<0.001} & \valinf{<0.001} & / & / & \valinf{0.002} \\
        Changing $L$ for $\chiczero$ and $\chictwo$ & \valinf{<0.001} & \valinf{0.003} & \valinf{<0.001} & \valinf{<0.001} & \valinf{0.002} \\
        New RBW radial parameter & \valinf{0.004} & \valinf{<0.001} & \valinf{<0.001} & \valinf{<0.001} & \valinf{<0.001} \\
        $10 \mev$ bin width & \valinf{<0.001} & \valinf{0.001} & \valinf{<0.001} & \valinf{<0.001} & \valinf{0.001} \\
        Changing momentum scaling & \valinf{0.015} & \valinf{0.018} & \valinf{0.018} & \valinf{0.019} & \valinf{0.019} \\
        \midrule
        Total systematic uncertainties & \valinf{0.015} & \valinf{0.018} & \valinf{0.018} & \valinf{0.021} & \valinf{0.021} \\
        \midrule
        Statistical uncertainties & \valinf{0.016} & \valinf{0.044} & \valinf{0.017} & \valinf{0.039} & \valinf{0.096} \\
        \bottomrule
    \end{tabular}
    \label{tab:syst_M}
\end{table}

\begin{table}
    \centering
    \caption{Relative systematic uncertainties on the charmonium-state widths, in percent.}
    \begin{tabular}{lccc}
        \toprule
        Source & $\Gamma_{\etacones}$ & $\Gamma_{\chiczero}$ & $\Gamma_{\etactwos}$ \\
        \midrule
        Single Gaussian resolution & \valinf{0.9} & \valinf{1.7} & \valinf{4.5} \\
        $\sigma_{\ccbar}/\sigma_{\etacones}$ from fit & \valinf{1.5} & \valinf{0.4} & \valinf{3.6} \\
        With $\chicone(3872)$ and $\chiczero(3915)$ & \valinf{<0.1} & \valinf{<0.1} & \valinf{0.7} \\
        New $\phi \phi$ mass fit background & \valinf{<0.1} & \valinf{0.3} & \valinf{<0.1} \\
        $M_\chicone$, $M_\chictwo$ at known values & \valinf{0.5} & \valinf{1.7} & \valinf{1.9} \\
        Changing $L$ for $\chiczero$ and $\chictwo$ & \valinf{<0.1} & \valinf{0.9} & \valinf{0.5} \\
        New RBW radial parameter & \valinf{<0.1} & \valinf{0.1} & \valinf{<0.1} \\
        $10 \mev$ bin width & \valinf{0.8} & \valinf{0.5} & \valinf{0.9} \\
        Changing momentum scaling & \valinf{<0.1} & \valinf{0.3} & \valinf{0.3} \\
        \midrule
        Total systematic uncertainties & \valinf{2.0} & \valinf{2.7} & \valinf{6.2} \\
        \midrule
        Statistical uncertainties & \valinf{3.2} & \!\!\!\valinf{17.8} & \!\!\!\valinf{27.4} \\
        \bottomrule
    \end{tabular}
    \label{tab:syst_G}
\end{table}

\begin{figure}
    \centering
    \includegraphics[width=1.0\linewidth]{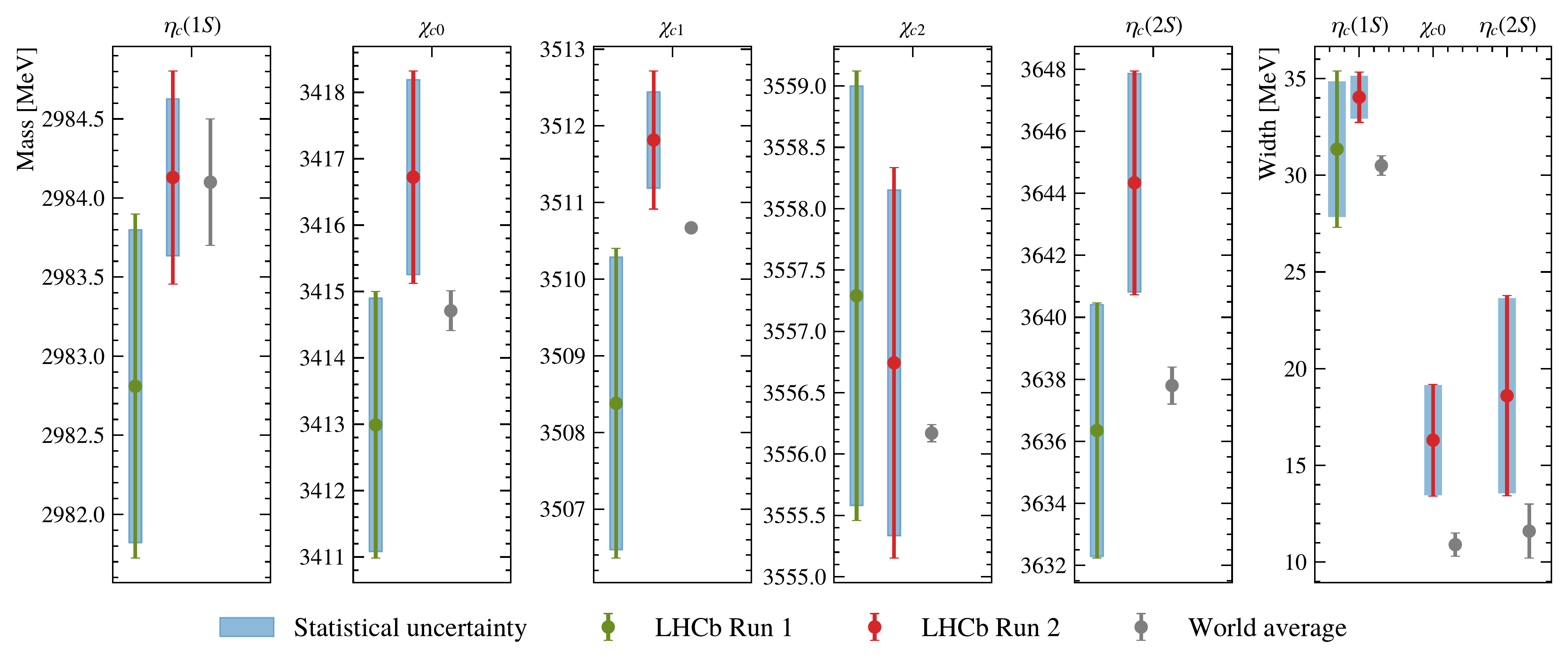}
    \caption{Masses and widths of the charmonium states measured in this analysis, in orange, with, from left to right: $M_{\etac(1S)}$, $M_{\chi_{c0}}$, $M_{\chi_{c1}}$, $M_{\chi_{c2}}$ and $M_{\etac(2S)}$. The rightmost plot shows $\Gamma_{\etac(1S)}$, $\Gamma_{\chi_{c0}}$ and $\Gamma_{\etac(2S)}$.  The corresponding world averages~\cite{PDG2024} are given in grey, and the Run~1 results~\cite{LHCb-PAPER-2017-007} in blue.}
    \label{fig:MG}
\end{figure}

The obtained masses and widths are given in Table~\ref{tab:MG} and Fig.~\ref{fig:MG}. All masses are consistent with the world averages within $2\sigma$. Along with the LHCb measurement using $\proton\antiproton$ in the final state~\cite{LHCb-PAPER-2019-024}, this is the most precise single measurement of $M_{\etacones}$. The measured widths are consistent with the world averages within $2.5\sigma$.

\begin{table}
    \centering
    \caption{Masses and widths of the charmonium states measured in this analysis compared to the world average values~\cite{PDG2024} and Run~1 results~\cite{LHCb-PAPER-2017-007}. The first uncertainty is statistical and the second systematic.}
    \begin{tabular}{lccc}
        \toprule
         & Run 1 & Run 2 & World average \\
         \midrule
        $M_{\etacones}$ [\!$\mev$] & $2982.8 \pm 1.0 \pm 0.5$ & $2984.1 \pm 0.5 \pm 0.5$ & $2984.1 \pm 0.4$ \\
        $M_{\chiczero}$ [\!$\mev$] & $3413.0 \pm 1.9 \pm 0.6$ & $3416.7 \pm 1.5 \pm 0.6$ & $3414.71 \pm 0.30$ \\
        $M_{\chicone}$ [\!$\mev$] & $3508.4 \pm 1.9 \pm 0.7$ & $3511.8 \pm 0.6 \pm 0.7$ & $3510.67 \pm 0.05$ \\
        $M_{\chictwo}$ [\!$\mev$] & $3557.3 \pm 1.7 \pm 0.7$ & $3556.7 \pm 1.4 \pm 0.7$ & $3556.17 \pm 0.07$ \\
        $M_{\etactwos}$ [\!$\mev$] & $3636.4 \pm 4.1 \pm 0.7$ & $3644.3 \pm 3.5 \pm 0.8$ & $3637.8~\, \pm 0.6~\,$ \\
        $\Gamma_{\etacones}$ [\!$\mev$] & $~~~31.4 \pm 3.5 \pm 2.0$ & $~~~34.0 \pm 1.1 \pm 0.7$ & $~~~30.5~\, \pm 0.5~\,$ \\
        $\Gamma_{\chiczero}$ [\!$\mev$] & --- & $~~~16.3 \pm 2.9 \pm 0.4$ & $~~~10.9~\, \pm 0.6~\,$ \\
        $\Gamma_{\etactwos}$ [\!$\mev$] & --- & $~~~18.6 \pm 5.1 \pm 1.2$ & $~~~11.6~\, \pm 1.4~\,$ \\
        \bottomrule
    \end{tabular}
    \label{tab:MG}
\end{table}

\section{Summary}
\label{sec:conc}

In summary, inclusive charmonia production in $b$-hadron decays is studied via their decays to a $\phi \phi$ pair, using data from LHCb Run~2 corresponding to an integrated luminosity of $5.9 \invfb$. The masses of the $\etacones$ and $\chicJ$ states are extracted, as well as the natural widths of the $\etacones$, $\chiczero$ and $\etactwos$ states. Interference effects are also included for the $\etacones$, $\chiczero$, $\chictwo$ and $\etactwos$ states. Using the known values of the $\etacones$ branching ratios, the branching ratios of inclusive production of $\chicJ$ states in $b$-hadron decays are determined to be
\begin{align*}
    &\BR (b \to \chiczero X) = (1.34 \pm 0.13 \pm 0.06 \pm 0.37) \times 10^{-3},\\
    &\BR (b \to \chicone X) = (1.58 \pm 0.12 \pm 0.09 \pm 0.44) \times 10^{-3},\\
    &\BR (b \to \chictwo X) = (0.55 \pm 0.08 \pm 0.05 \pm 0.15) \times 10^{-3}.
\end{align*}
These measurements are limited by the knowledge of the normalisation channel. Nevertheless, they constitute the most precise measurements of the branching fractions from an admixture of $\Bz$, $\Bp$, $\Bs$, $\Bc$ mesons and $b$-baryons for $\chictwo$ production, and provide useful contributions for $\chiczero$ and $\chicone$ production. The branching-fraction ratios for the $\chic_{1,2}$ states with $\chiczero$ are determined to be
\begin{align*}
    &\frac{\BR (b \to \chicone X)}{\BR (b \to \chiczero X)} = 1.18 \pm 0.13 \pm 0.08 \pm 0.07,\\
    &\frac{\BR (b \to \chictwo X)}{\BR (b \to \chiczero X)} = 0.41 \pm 0.07 \pm 0.04 \pm 0.03.
\end{align*}
The production of the $\etactwos$ state in $b$-hadron decays is determined to be
\begin{equation*}
    \BR (b \to \etactwos X) \times \BR (\etactwos \to \phi \phi) = (4.0 \pm 0.6 \pm 0.6 \pm 1.1) \times 10^{-7}.
\end{equation*}
All measurements are compatible with previous results based on LHCb Run~1 data~\cite{LHCb-PAPER-2017-007}. The $\etacones$ mass is determined to be \mbox{$M_{\etacones} = 2984.1\pm 0.5 \pm 0.5$\mev}, representing the most precise measurement to date, with equal precision to that measured from $\etacones \to p\overline{p}$ decays at \lhcb~\cite{LHCb-PAPER-2019-024}.

\section*{Acknowledgements}
%
%
\noindent We thank Feng-Kun Guo and Alexander Bondar for our discussions concerning the possible interference effects. We express our gratitude to our colleagues in the CERN
accelerator departments for the excellent performance of the LHC. We
thank the technical and administrative staff at the LHCb
institutes.
We acknowledge support from CERN and from the national agencies:
ARC (Australia);
CAPES, CNPq, FAPERJ and FINEP (Brazil); 
MOST and NSFC (China); 
CNRS/IN2P3 and CEA (France);  
BMFTR, DFG and MPG (Germany);
INFN (Italy); 
NWO (Netherlands); 
MNiSW and NCN (Poland); 
MCID/IFA (Romania); 
MICIU and AEI (Spain);
SNSF and SER (Switzerland); 
NASU (Ukraine); 
STFC (United Kingdom); 
DOE NP and NSF (USA).
We acknowledge the computing resources that are provided by ARDC (Australia), 
CBPF (Brazil),
CERN, 
IHEP and LZU (China),
IN2P3 (France), 
KIT and DESY (Germany), 
INFN (Italy), 
SURF (Netherlands),
Polish WLCG (Poland),
IFIN-HH (Romania), 
PIC (Spain), CSCS (Switzerland), 
GridPP (United Kingdom),
and NSF (USA).  
We are indebted to the communities behind the multiple open-source
software packages on which we depend.
Individual groups or members have received support from
Key Research Program of Frontier Sciences of CAS, CAS PIFI, CAS CCEPP (China); 
Minciencias (Colombia);
EPLANET, Marie Sk\l{}odowska-Curie Actions, ERC and NextGenerationEU (European Union);
A*MIDEX, ANR, IPhU and Labex P2IO, and R\'{e}gion Auvergne-Rh\^{o}ne-Alpes (France);
Alexander-von-Humboldt Foundation (Germany);
ICSC (Italy); 
Severo Ochoa and Mar\'ia de Maeztu Units of Excellence, GVA, XuntaGal, GENCAT, InTalent-Inditex and Prog.~Atracci\'on Talento CM (Spain);
the Leverhulme Trust, the Royal Society and UKRI (United Kingdom).

\clearpage
\addcontentsline{toc}{section}{References}
\bibliographystyle{LHCb/LHCb}
\bibliography{main,LHCb/standard,LHCb/LHCb-PAPER,LHCb/LHCb-CONF,LHCb/LHCb-DP,LHCb/LHCb-TDR}

\ifx\mcitethebibliography\mciteundefinedmacro
\PackageError{LHCb.bst}{mciteplus.sty has not been loaded}
{This bibstyle requires the use of the mciteplus package.}\fi
\providecommand{\href}[2]{#2}
\begin{mcitethebibliography}{10}
\mciteSetBstSublistMode{n}
\mciteSetBstMaxWidthForm{subitem}{\alph{mcitesubitemcount})}
\mciteSetBstSublistLabelBeginEnd{\mcitemaxwidthsubitemform\space}
{\relax}{\relax}

\bibitem{PhysRevD.51.1125}
G.~T. Bodwin, E.~Braaten, and G.~P. Lepage, \ifthenelse{\boolean{articletitles}}{\emph{{Rigorous QCD analysis of inclusive annihilation and production of heavy quarkonium}}, }{}\href{https://doi.org/10.1103/PhysRevD.51.1125}{Phys.\ Rev.\  \textbf{D51} (1995) 1125}, Erratum \href{https://doi.org/10.1103/PhysRevD.55.5853}{ibid.\   \textbf{D55} (1997) 5853}, \href{http://arxiv.org/abs/hep-ph/9407339}{{\normalfont\ttfamily arXiv:hep-ph/9407339}}\relax
\mciteBstWouldAddEndPuncttrue
\mciteSetBstMidEndSepPunct{\mcitedefaultmidpunct}
{\mcitedefaultendpunct}{\mcitedefaultseppunct}\relax
\EndOfBibitem
\bibitem{Beneke:1998ks}
M.~Beneke, F.~Maltoni, and I.~Z. Rothstein, \ifthenelse{\boolean{articletitles}}{\emph{{{QCD} analysis of inclusive {B} decay into charmonium}}, }{}\href{https://doi.org/10.1103/PhysRevD.59.054003}{Phys.\ Rev.\  \textbf{D59} (1999) 054003}, \href{http://arxiv.org/abs/hep-ph/9808360}{{\normalfont\ttfamily arXiv:hep-ph/9808360}}\relax
\mciteBstWouldAddEndPuncttrue
\mciteSetBstMidEndSepPunct{\mcitedefaultmidpunct}
{\mcitedefaultendpunct}{\mcitedefaultseppunct}\relax
\EndOfBibitem
\bibitem{Beneke_2000}
M.~Beneke, G.~A. Schuler, and S.~Wolf, \ifthenelse{\boolean{articletitles}}{\emph{{Quarkonium momentum distributions in photoproduction and B decay}}, }{}\href{https://doi.org/10.1103/PhysRevD.62.034004}{Phys.\ Rev.\  \textbf{D62} (2000) 034004}, \href{http://arxiv.org/abs/hep-ph/0001062}{{\normalfont\ttfamily arXiv:hep-ph/0001062}}\relax
\mciteBstWouldAddEndPuncttrue
\mciteSetBstMidEndSepPunct{\mcitedefaultmidpunct}
{\mcitedefaultendpunct}{\mcitedefaultseppunct}\relax
\EndOfBibitem
\bibitem{Burns_2011}
T.~J. Burns {\em et~al.}, \ifthenelse{\boolean{articletitles}}{\emph{{Momentum distribution of \jpsi in B decays}}, }{}\href{https://doi.org/10.1103/PhysRevD.83.114029}{Phys.\ Rev.\  \textbf{D83} (2011) 114029}, \href{http://arxiv.org/abs/1104.1781}{{\normalfont\ttfamily arXiv:1104.1781}}\relax
\mciteBstWouldAddEndPuncttrue
\mciteSetBstMidEndSepPunct{\mcitedefaultmidpunct}
{\mcitedefaultendpunct}{\mcitedefaultseppunct}\relax
\EndOfBibitem
\bibitem{Schuler_1999}
G.~A. Schuler, \ifthenelse{\boolean{articletitles}}{\emph{{Testing factorization of charmonium production}}, }{}\href{https://doi.org/10.1007/s100529900948}{Eur.\ Phys.\ J.\  \textbf{C8} (1999) 273}, \href{http://arxiv.org/abs/hep-ph/9804349}{{\normalfont\ttfamily arXiv:hep-ph/9804349}}\relax
\mciteBstWouldAddEndPuncttrue
\mciteSetBstMidEndSepPunct{\mcitedefaultmidpunct}
{\mcitedefaultendpunct}{\mcitedefaultseppunct}\relax
\EndOfBibitem
\bibitem{LHCb-PAPER-2011-003}
LHCb collaboration, R.~Aaij {\em et~al.}, \ifthenelse{\boolean{articletitles}}{\emph{{Measurement of \jpsi production in \proton\proton collisions at \mbox{$\sqs=7$\tev}}}, }{}\href{https://doi.org/10.1140/epjc/s10052-011-1645-y}{Eur.\ Phys.\ J.\  \textbf{C71} (2011) 1645}, \href{http://arxiv.org/abs/1103.0423}{{\normalfont\ttfamily arXiv:1103.0423}}\relax
\mciteBstWouldAddEndPuncttrue
\mciteSetBstMidEndSepPunct{\mcitedefaultmidpunct}
{\mcitedefaultendpunct}{\mcitedefaultseppunct}\relax
\EndOfBibitem
\bibitem{LHCB-PAPER-2015-037}
LHCb collaboration, R.~Aaij {\em et~al.}, \ifthenelse{\boolean{articletitles}}{\emph{{Measurement of forward \jpsi production cross-sections in \proton\proton collisions at \mbox{$\sqs=13$\tev}}}, }{}\href{https://doi.org/10.1007/JHEP10(2015)172}{JHEP \textbf{10} (2015) 172}, Erratum \href{https://doi.org/10.1007/JHEP05(2017)063}{ibid.\   \textbf{05} (2017) 063}, \href{http://arxiv.org/abs/1509.00771}{{\normalfont\ttfamily arXiv:1509.00771}}\relax
\mciteBstWouldAddEndPuncttrue
\mciteSetBstMidEndSepPunct{\mcitedefaultmidpunct}
{\mcitedefaultendpunct}{\mcitedefaultseppunct}\relax
\EndOfBibitem
\bibitem{LHCb-PAPER-2018-049}
LHCb collaboration, R.~Aaij {\em et~al.}, \ifthenelse{\boolean{articletitles}}{\emph{{Measurement of \psitwos production cross-sections in proton-proton collisions at $\sqs = 7$ and $13$\tev}}, }{}\href{https://doi.org/10.1140/epjc/s10052-020-7638-y}{Eur.\ Phys.\ J.\  \textbf{C80} (2020) 185}, \href{http://arxiv.org/abs/1908.03099}{{\normalfont\ttfamily arXiv:1908.03099}}\relax
\mciteBstWouldAddEndPuncttrue
\mciteSetBstMidEndSepPunct{\mcitedefaultmidpunct}
{\mcitedefaultendpunct}{\mcitedefaultseppunct}\relax
\EndOfBibitem
\bibitem{LHCb-PAPER-2017-007}
LHCb collaboration, R.~Aaij {\em et~al.}, \ifthenelse{\boolean{articletitles}}{\emph{{Study of charmonium production in $b$-hadron decays and first evidence for the decay \mbox{\decay{\Bs}{\phi\phi\phi}}}}, }{}\href{https://doi.org/10.1140/epjc/s10052-017-5151-8}{Eur.\ Phys.\ J.\  \textbf{C77} (2017) 609}, \href{http://arxiv.org/abs/1706.07013}{{\normalfont\ttfamily arXiv:1706.07013}}\relax
\mciteBstWouldAddEndPuncttrue
\mciteSetBstMidEndSepPunct{\mcitedefaultmidpunct}
{\mcitedefaultendpunct}{\mcitedefaultseppunct}\relax
\EndOfBibitem
\bibitem{LHCb-PAPER-2019-024}
LHCb collaboration, R.~Aaij {\em et~al.}, \ifthenelse{\boolean{articletitles}}{\emph{{Measurement of the $\eta_c(1S)$ production cross-section in $pp$ collisions at $\sqrt{s}=13$\tev}}, }{}\href{https://doi.org/10.1140/epjc/s10052-020-7733-0}{Eur.\ Phys.\ J.\  \textbf{C80} (2020) 191}, \href{http://arxiv.org/abs/1911.03326}{{\normalfont\ttfamily arXiv:1911.03326}}\relax
\mciteBstWouldAddEndPuncttrue
\mciteSetBstMidEndSepPunct{\mcitedefaultmidpunct}
{\mcitedefaultendpunct}{\mcitedefaultseppunct}\relax
\EndOfBibitem
\bibitem{LHCb-PAPER-2024-004}
LHCb collaboration, R.~Aaij {\em et~al.}, \ifthenelse{\boolean{articletitles}}{\emph{{Study of charmonium production via the decay to $p\bar p$ at $\sqrt{s} = 13$\tev}}, }{}\href{https://doi.org/10.1140/epjc/s10052-024-13478-y}{{Eur.\ Phys.\ J.\ } \textbf{C84} (2024) 1274}, \href{http://arxiv.org/abs/2407.14261}{{\normalfont\ttfamily arXiv:2407.14261}}\relax
\mciteBstWouldAddEndPuncttrue
\mciteSetBstMidEndSepPunct{\mcitedefaultmidpunct}
{\mcitedefaultendpunct}{\mcitedefaultseppunct}\relax
\EndOfBibitem
\bibitem{LHCb-PAPER-2016-016}
LHCb collaboration, R.~Aaij {\em et~al.}, \ifthenelse{\boolean{articletitles}}{\emph{{Observation of \mbox{\decay{\etac(2S)}{\proton\antiproton}} and search for \mbox{\decay{X(3872)}{\proton\antiproton}} decays}}, }{}\href{https://doi.org/10.1016/j.physletb.2017.03.046}{Phys.\ Lett.\  \textbf{B769} (2017) 305}, \href{http://arxiv.org/abs/1607.06446}{{\normalfont\ttfamily arXiv:1607.06446}}\relax
\mciteBstWouldAddEndPuncttrue
\mciteSetBstMidEndSepPunct{\mcitedefaultmidpunct}
{\mcitedefaultendpunct}{\mcitedefaultseppunct}\relax
\EndOfBibitem
\bibitem{LHCb-PAPER-2022-051}
LHCb collaboration, R.~Aaij {\em et~al.}, \ifthenelse{\boolean{articletitles}}{\emph{{Study of charmonium decays to $\KS\kaon\pion$ in the $\B \to (\KS \kaon\pion)\kaon$ channels}}, }{}\href{https://doi.org/10.1103/PhysRevD.108.032010}{Phys.\ Rev.\  \textbf{D108} (2023) 032010}, \href{http://arxiv.org/abs/2304.14891}{{\normalfont\ttfamily arXiv:2304.14891}}\relax
\mciteBstWouldAddEndPuncttrue
\mciteSetBstMidEndSepPunct{\mcitedefaultmidpunct}
{\mcitedefaultendpunct}{\mcitedefaultseppunct}\relax
\EndOfBibitem
\bibitem{BESIII:2011ab}
BESIII collaboration, M.~Ablikim {\em et~al.}, \ifthenelse{\boolean{articletitles}}{\emph{{Measurements of the mass and width of the $\eta_c$ using the decay $\psi' \to \gamma \eta_c$}}, }{}\href{https://doi.org/10.1103/PhysRevLett.108.222002}{Phys.\ Rev.\ Lett.\  \textbf{108} (2012) 222002}, \href{http://arxiv.org/abs/1111.0398}{{\normalfont\ttfamily arXiv:1111.0398}}\relax
\mciteBstWouldAddEndPuncttrue
\mciteSetBstMidEndSepPunct{\mcitedefaultmidpunct}
{\mcitedefaultendpunct}{\mcitedefaultseppunct}\relax
\EndOfBibitem
\bibitem{BESIII:2012urf}
BESIII collaboration, M.~Ablikim {\em et~al.}, \ifthenelse{\boolean{articletitles}}{\emph{{Study of $\psi(3686)\to\pi^0 h_c, h_c\to\gamma\eta_c$ via $\eta_c$ exclusive decays}}, }{}\href{https://doi.org/10.1103/PhysRevD.86.092009}{Phys.\ Rev.\  \textbf{D86} (2012) 092009}, \href{http://arxiv.org/abs/1209.4963}{{\normalfont\ttfamily arXiv:1209.4963}}\relax
\mciteBstWouldAddEndPuncttrue
\mciteSetBstMidEndSepPunct{\mcitedefaultmidpunct}
{\mcitedefaultendpunct}{\mcitedefaultseppunct}\relax
\EndOfBibitem
\bibitem{BaBar:2011gos}
BaBar collaboration, P.~del Amo~Sanchez {\em et~al.}, \ifthenelse{\boolean{articletitles}}{\emph{{Observation of $\eta_c(1S)$ and $\eta_c(2S)$ decays to $K^+ K^- \pi^+ \pi^- \pi^0$ in two-photon interactions}}, }{}\href{https://doi.org/10.1103/PhysRevD.84.012004}{Phys.\ Rev.\  \textbf{D84} (2011) 012004}, \href{http://arxiv.org/abs/1103.3971}{{\normalfont\ttfamily arXiv:1103.3971}}\relax
\mciteBstWouldAddEndPuncttrue
\mciteSetBstMidEndSepPunct{\mcitedefaultmidpunct}
{\mcitedefaultendpunct}{\mcitedefaultseppunct}\relax
\EndOfBibitem
\bibitem{BESIII:2023lcc}
BESIII collaboration, M.~Ablikim {\em et~al.}, \ifthenelse{\boolean{articletitles}}{\emph{{Updated measurements of the M1 transition $\psi(3686)\to\gamma\eta_c(2S)$ with $\eta_c(2S)\to K\Kbar\pi$}}, }{}\href{https://doi.org/10.1103/PhysRevD.109.032004}{Phys.\ Rev.\  \textbf{D109} (2024) 032004}, \href{http://arxiv.org/abs/2309.14689}{{\normalfont\ttfamily arXiv:2309.14689}}\relax
\mciteBstWouldAddEndPuncttrue
\mciteSetBstMidEndSepPunct{\mcitedefaultmidpunct}
{\mcitedefaultendpunct}{\mcitedefaultseppunct}\relax
\EndOfBibitem
\bibitem{BES:1998ser}
BES collaboration, J.~Z. Bai {\em et~al.}, \ifthenelse{\boolean{articletitles}}{\emph{{Study of the hadronic decays of \chic states}}, }{}\href{https://doi.org/10.1103/PhysRevD.60.072001}{Phys.\ Rev.\  \textbf{D60} (1999) 072001}, \href{http://arxiv.org/abs/hep-ex/9812016}{{\normalfont\ttfamily arXiv:hep-ex/9812016}}\relax
\mciteBstWouldAddEndPuncttrue
\mciteSetBstMidEndSepPunct{\mcitedefaultmidpunct}
{\mcitedefaultendpunct}{\mcitedefaultseppunct}\relax
\EndOfBibitem
\bibitem{BES:2005bmx}
BES collaboration, M.~Ablikim {\em et~al.}, \ifthenelse{\boolean{articletitles}}{\emph{{Precise measurement of spin-averaged $\chicJ(1P)$ mass using photon conversions in $\psitwos \to \gamma \chicJ$}}, }{}\href{https://doi.org/10.1103/PhysRevD.71.092002}{Phys.\ Rev.\  \textbf{D71} (2005) 092002}, \href{http://arxiv.org/abs/hep-ex/0502031}{{\normalfont\ttfamily arXiv:hep-ex/0502031}}\relax
\mciteBstWouldAddEndPuncttrue
\mciteSetBstMidEndSepPunct{\mcitedefaultmidpunct}
{\mcitedefaultendpunct}{\mcitedefaultseppunct}\relax
\EndOfBibitem
\bibitem{FermilabE835:2002pkx}
E835 collaboration, S.~Bagnasco {\em et~al.}, \ifthenelse{\boolean{articletitles}}{\emph{{New measurements of the resonance parameters of the $\chiczero(1^3 P_0)$ state of charmonium}}, }{}\href{https://doi.org/10.1016/S0370-2693(02)01657-X}{Phys.\ Lett.\  \textbf{B533} (2002) 237}\relax
\mciteBstWouldAddEndPuncttrue
\mciteSetBstMidEndSepPunct{\mcitedefaultmidpunct}
{\mcitedefaultendpunct}{\mcitedefaultseppunct}\relax
\EndOfBibitem
\bibitem{FermilabE835:2003gwi}
E835 collaboration, M.~Andreotti {\em et~al.}, \ifthenelse{\boolean{articletitles}}{\emph{{Interference study of the $\chi_{c0}(1^{3}P_0)$ in the reaction $\proton\antiproton \to \piz \piz$}}, }{}\href{https://doi.org/10.1103/PhysRevLett.91.091801}{Phys.\ Rev.\ Lett.\  \textbf{91} (2003) 091801}, \href{http://arxiv.org/abs/hep-ex/0308055}{{\normalfont\ttfamily arXiv:hep-ex/0308055}}\relax
\mciteBstWouldAddEndPuncttrue
\mciteSetBstMidEndSepPunct{\mcitedefaultmidpunct}
{\mcitedefaultendpunct}{\mcitedefaultseppunct}\relax
\EndOfBibitem
\bibitem{E760:1991qim}
E760 collaboration, T.~A. Armstrong {\em et~al.}, \ifthenelse{\boolean{articletitles}}{\emph{{Study of the $\chi_1$ and $\chi_2$ charmonium states formed in $\antiproton \proton$ annihilations}}, }{}\href{https://doi.org/10.1016/0550-3213(92)90448-K}{Nucl.\ Phys.\  \textbf{B373} (1992) 35}\relax
\mciteBstWouldAddEndPuncttrue
\mciteSetBstMidEndSepPunct{\mcitedefaultmidpunct}
{\mcitedefaultendpunct}{\mcitedefaultseppunct}\relax
\EndOfBibitem
\bibitem{Andreotti:2005ts}
E835 collaboration, M.~Andreotti {\em et~al.}, \ifthenelse{\boolean{articletitles}}{\emph{{Measurement of the resonance parameters of the $\chi_1(1^3P_1)$ and $\chi_2(1^3P_2)$ states of charmonium formed in antiproton-proton annihilations}}, }{}\href{https://doi.org/10.1016/j.nuclphysb.2005.03.042}{Nucl.\ Phys.\  \textbf{B717} (2005) 34}, \href{http://arxiv.org/abs/hep-ex/0503022}{{\normalfont\ttfamily arXiv:hep-ex/0503022}}\relax
\mciteBstWouldAddEndPuncttrue
\mciteSetBstMidEndSepPunct{\mcitedefaultmidpunct}
{\mcitedefaultendpunct}{\mcitedefaultseppunct}\relax
\EndOfBibitem
\bibitem{LHCb-PAPER-2017-036}
LHCb collaboration, R.~Aaij {\em et~al.}, \ifthenelse{\boolean{articletitles}}{\emph{{$\chi_{c1}$ and $\chi_{c2}$ resonance parameters with the decays \mbox{\decay{\chi_{c1,c2}}{\jpsi\mumu}}}}, }{}\href{https://doi.org/10.1103/PhysRevLett.119.221801}{Phys.\ Rev.\ Lett.\  \textbf{119} (2017) 221801}, \href{http://arxiv.org/abs/1709.04247}{{\normalfont\ttfamily arXiv:1709.04247}}\relax
\mciteBstWouldAddEndPuncttrue
\mciteSetBstMidEndSepPunct{\mcitedefaultmidpunct}
{\mcitedefaultendpunct}{\mcitedefaultseppunct}\relax
\EndOfBibitem
\bibitem{LHCb-DP-2008-001}
LHCb collaboration, A.~A. Alves~Jr.\ {\em et~al.}, \ifthenelse{\boolean{articletitles}}{\emph{{The \lhcb detector at the LHC}}, }{}\href{https://doi.org/10.1088/1748-0221/3/08/S08005}{JINST \textbf{3} (2008) S08005}\relax
\mciteBstWouldAddEndPuncttrue
\mciteSetBstMidEndSepPunct{\mcitedefaultmidpunct}
{\mcitedefaultendpunct}{\mcitedefaultseppunct}\relax
\EndOfBibitem
\bibitem{LHCb-DP-2014-002}
LHCb collaboration, R.~Aaij {\em et~al.}, \ifthenelse{\boolean{articletitles}}{\emph{{LHCb detector performance}}, }{}\href{https://doi.org/10.1142/S0217751X15300227}{Int.\ J.\ Mod.\ Phys.\  \textbf{A30} (2015) 1530022}, \href{http://arxiv.org/abs/1412.6352}{{\normalfont\ttfamily arXiv:1412.6352}}\relax
\mciteBstWouldAddEndPuncttrue
\mciteSetBstMidEndSepPunct{\mcitedefaultmidpunct}
{\mcitedefaultendpunct}{\mcitedefaultseppunct}\relax
\EndOfBibitem
\bibitem{LHCb-DP-2014-001}
R.~Aaij {\em et~al.}, \ifthenelse{\boolean{articletitles}}{\emph{{Performance of the LHCb Vertex Locator}}, }{}\href{https://doi.org/10.1088/1748-0221/9/09/P09007}{JINST \textbf{9} (2014) P09007}, \href{http://arxiv.org/abs/1405.7808}{{\normalfont\ttfamily arXiv:1405.7808}}\relax
\mciteBstWouldAddEndPuncttrue
\mciteSetBstMidEndSepPunct{\mcitedefaultmidpunct}
{\mcitedefaultendpunct}{\mcitedefaultseppunct}\relax
\EndOfBibitem
\bibitem{LHCb-DP-2017-001}
P.~d'Argent {\em et~al.}, \ifthenelse{\boolean{articletitles}}{\emph{{Improved performance of the LHCb Outer Tracker in LHC Run 2}}, }{}\href{https://doi.org/10.1088/1748-0221/12/11/P11016}{JINST \textbf{12} (2017) P11016}, \href{http://arxiv.org/abs/1708.00819}{{\normalfont\ttfamily arXiv:1708.00819}}\relax
\mciteBstWouldAddEndPuncttrue
\mciteSetBstMidEndSepPunct{\mcitedefaultmidpunct}
{\mcitedefaultendpunct}{\mcitedefaultseppunct}\relax
\EndOfBibitem
\bibitem{LHCb-DP-2012-003}
M.~Adinolfi {\em et~al.}, \ifthenelse{\boolean{articletitles}}{\emph{{Performance of the \lhcb RICH detector at the LHC}}, }{}\href{https://doi.org/10.1140/epjc/s10052-013-2431-9}{Eur.\ Phys.\ J.\  \textbf{C73} (2013) 2431}, \href{http://arxiv.org/abs/1211.6759}{{\normalfont\ttfamily arXiv:1211.6759}}\relax
\mciteBstWouldAddEndPuncttrue
\mciteSetBstMidEndSepPunct{\mcitedefaultmidpunct}
{\mcitedefaultendpunct}{\mcitedefaultseppunct}\relax
\EndOfBibitem
\bibitem{LHCb-DP-2012-002}
A.~A. Alves~Jr.\ {\em et~al.}, \ifthenelse{\boolean{articletitles}}{\emph{{Performance of the LHCb muon system}}, }{}\href{https://doi.org/10.1088/1748-0221/8/02/P02022}{JINST \textbf{8} (2013) P02022}, \href{http://arxiv.org/abs/1211.1346}{{\normalfont\ttfamily arXiv:1211.1346}}\relax
\mciteBstWouldAddEndPuncttrue
\mciteSetBstMidEndSepPunct{\mcitedefaultmidpunct}
{\mcitedefaultendpunct}{\mcitedefaultseppunct}\relax
\EndOfBibitem
\bibitem{LHCb-DP-2012-004}
R.~Aaij {\em et~al.}, \ifthenelse{\boolean{articletitles}}{\emph{{The \lhcb trigger and its performance in 2011}}, }{}\href{https://doi.org/10.1088/1748-0221/8/04/P04022}{JINST \textbf{8} (2013) P04022}, \href{http://arxiv.org/abs/1211.3055}{{\normalfont\ttfamily arXiv:1211.3055}}\relax
\mciteBstWouldAddEndPuncttrue
\mciteSetBstMidEndSepPunct{\mcitedefaultmidpunct}
{\mcitedefaultendpunct}{\mcitedefaultseppunct}\relax
\EndOfBibitem
\bibitem{BBDT}
V.~V. Gligorov and M.~Williams, \ifthenelse{\boolean{articletitles}}{\emph{{Efficient, reliable and fast high-level triggering using a bonsai boosted decision tree}}, }{}\href{https://doi.org/10.1088/1748-0221/8/02/P02013}{JINST \textbf{8} (2013) P02013}, \href{http://arxiv.org/abs/1210.6861}{{\normalfont\ttfamily arXiv:1210.6861}}\relax
\mciteBstWouldAddEndPuncttrue
\mciteSetBstMidEndSepPunct{\mcitedefaultmidpunct}
{\mcitedefaultendpunct}{\mcitedefaultseppunct}\relax
\EndOfBibitem
\bibitem{LHCb-PROC-2015-018}
T.~Likhomanenko {\em et~al.}, \ifthenelse{\boolean{articletitles}}{\emph{{LHCb topological trigger reoptimization}}, }{}\href{https://doi.org/10.1088/1742-6596/664/8/082025}{J.\ Phys.\ Conf.\ Ser.\  \textbf{664} (2015) 082025}, \href{http://arxiv.org/abs/1510.00572}{{\normalfont\ttfamily arXiv:1510.00572}}\relax
\mciteBstWouldAddEndPuncttrue
\mciteSetBstMidEndSepPunct{\mcitedefaultmidpunct}
{\mcitedefaultendpunct}{\mcitedefaultseppunct}\relax
\EndOfBibitem
\bibitem{PDG2024}
Particle Data Group, S.~Navas {\em et~al.}, \ifthenelse{\boolean{articletitles}}{\emph{{\href{http://pdg.lbl.gov/}{Review of particle physics}}}, }{}\href{https://doi.org/10.1103/PhysRevD.110.030001}{Phys.\ Rev.\  \textbf{D110} (2024) 030001}\relax
\mciteBstWouldAddEndPuncttrue
\mciteSetBstMidEndSepPunct{\mcitedefaultmidpunct}
{\mcitedefaultendpunct}{\mcitedefaultseppunct}\relax
\EndOfBibitem
\bibitem{Stripping}
N.~Grieser {\em et~al.}, \ifthenelse{\boolean{articletitles}}{\emph{{The LHCb stripping project: Sustainable legacy data processing for high-energy physics}}, }{}\href{https://doi.org/10.1007/s41781-025-00151-6}{Comput.\ Softw.\ Big.\ Sci.\  \textbf{9} (2025) 21}, \href{http://arxiv.org/abs/2509.05294}{{\normalfont\ttfamily arXiv:2509.05294}}\relax
\mciteBstWouldAddEndPuncttrue
\mciteSetBstMidEndSepPunct{\mcitedefaultmidpunct}
{\mcitedefaultendpunct}{\mcitedefaultseppunct}\relax
\EndOfBibitem
\bibitem{LHCb-PAPER-2012-048}
LHCb collaboration, R.~Aaij {\em et~al.}, \ifthenelse{\boolean{articletitles}}{\emph{{Measurements of the \Lb, \Xibm, and \Omegab baryon masses}}, }{}\href{https://doi.org/10.1103/PhysRevLett.110.182001}{Phys.\ Rev.\ Lett.\  \textbf{110} (2013) 182001}, \href{http://arxiv.org/abs/1302.1072}{{\normalfont\ttfamily arXiv:1302.1072}}\relax
\mciteBstWouldAddEndPuncttrue
\mciteSetBstMidEndSepPunct{\mcitedefaultmidpunct}
{\mcitedefaultendpunct}{\mcitedefaultseppunct}\relax
\EndOfBibitem
\bibitem{LHCb-PAPER-2013-011}
LHCb collaboration, R.~Aaij {\em et~al.}, \ifthenelse{\boolean{articletitles}}{\emph{{Precision measurement of \D meson mass differences}}, }{}\href{https://doi.org/10.1007/JHEP06(2013)065}{JHEP \textbf{06} (2013) 065}, \href{http://arxiv.org/abs/1304.6865}{{\normalfont\ttfamily arXiv:1304.6865}}\relax
\mciteBstWouldAddEndPuncttrue
\mciteSetBstMidEndSepPunct{\mcitedefaultmidpunct}
{\mcitedefaultendpunct}{\mcitedefaultseppunct}\relax
\EndOfBibitem
\bibitem{Sjostrand:2007gs}
T.~Sj\"{o}strand, S.~Mrenna, and P.~Skands, \ifthenelse{\boolean{articletitles}}{\emph{{A brief introduction to PYTHIA 8.1}}, }{}\href{https://doi.org/10.1016/j.cpc.2008.01.036}{Comput.\ Phys.\ Commun.\  \textbf{178} (2008) 852}, \href{http://arxiv.org/abs/0710.3820}{{\normalfont\ttfamily arXiv:0710.3820}}\relax
\mciteBstWouldAddEndPuncttrue
\mciteSetBstMidEndSepPunct{\mcitedefaultmidpunct}
{\mcitedefaultendpunct}{\mcitedefaultseppunct}\relax
\EndOfBibitem
\bibitem{Sjostrand:2006za}
T.~Sj\"{o}strand, S.~Mrenna, and P.~Skands, \ifthenelse{\boolean{articletitles}}{\emph{{PYTHIA 6.4 physics and manual}}, }{}\href{https://doi.org/10.1088/1126-6708/2006/05/026}{JHEP \textbf{05} (2006) 026}, \href{http://arxiv.org/abs/hep-ph/0603175}{{\normalfont\ttfamily arXiv:hep-ph/0603175}}\relax
\mciteBstWouldAddEndPuncttrue
\mciteSetBstMidEndSepPunct{\mcitedefaultmidpunct}
{\mcitedefaultendpunct}{\mcitedefaultseppunct}\relax
\EndOfBibitem
\bibitem{LHCb-PROC-2010-056}
I.~Belyaev {\em et~al.}, \ifthenelse{\boolean{articletitles}}{\emph{{Handling of the generation of primary events in Gauss, the LHCb simulation framework}}, }{}\href{https://doi.org/10.1088/1742-6596/331/3/032047}{J.\ Phys.\ Conf.\ Ser.\  \textbf{331} (2011) 032047}\relax
\mciteBstWouldAddEndPuncttrue
\mciteSetBstMidEndSepPunct{\mcitedefaultmidpunct}
{\mcitedefaultendpunct}{\mcitedefaultseppunct}\relax
\EndOfBibitem
\bibitem{Lange:2001uf}
D.~J. Lange, \ifthenelse{\boolean{articletitles}}{\emph{{The EvtGen particle decay simulation package}}, }{}\href{https://doi.org/10.1016/S0168-9002(01)00089-4}{Nucl.\ Instrum.\ Meth.\  \textbf{A462} (2001) 152}\relax
\mciteBstWouldAddEndPuncttrue
\mciteSetBstMidEndSepPunct{\mcitedefaultmidpunct}
{\mcitedefaultendpunct}{\mcitedefaultseppunct}\relax
\EndOfBibitem
\bibitem{davidson2015photos}
N.~Davidson, T.~Przedzinski, and Z.~Was, \ifthenelse{\boolean{articletitles}}{\emph{{PHOTOS interface in C++: Technical and physics documentation}}, }{}\href{https://doi.org/https://doi.org/10.1016/j.cpc.2015.09.013}{Comput.\ Phys.\ Commun.\  \textbf{199} (2016) 86}, \href{http://arxiv.org/abs/1011.0937}{{\normalfont\ttfamily arXiv:1011.0937}}\relax
\mciteBstWouldAddEndPuncttrue
\mciteSetBstMidEndSepPunct{\mcitedefaultmidpunct}
{\mcitedefaultendpunct}{\mcitedefaultseppunct}\relax
\EndOfBibitem
\bibitem{Allison:2006ve}
Geant4 collaboration, J.~Allison {\em et~al.}, \ifthenelse{\boolean{articletitles}}{\emph{{Geant4 developments and applications}}, }{}\href{https://doi.org/10.1109/TNS.2006.869826}{IEEE Trans.\ Nucl.\ Sci.\  \textbf{53} (2006) 270}\relax
\mciteBstWouldAddEndPuncttrue
\mciteSetBstMidEndSepPunct{\mcitedefaultmidpunct}
{\mcitedefaultendpunct}{\mcitedefaultseppunct}\relax
\EndOfBibitem
\bibitem{Agostinelli:2002hh}
Geant4 collaboration, S.~Agostinelli {\em et~al.}, \ifthenelse{\boolean{articletitles}}{\emph{{Geant4: A simulation toolkit}}, }{}\href{https://doi.org/10.1016/S0168-9002(03)01368-8}{Nucl.\ Instrum.\ Meth.\  \textbf{A506} (2003) 250}\relax
\mciteBstWouldAddEndPuncttrue
\mciteSetBstMidEndSepPunct{\mcitedefaultmidpunct}
{\mcitedefaultendpunct}{\mcitedefaultseppunct}\relax
\EndOfBibitem
\bibitem{LHCb-PROC-2011-006}
M.~Clemencic {\em et~al.}, \ifthenelse{\boolean{articletitles}}{\emph{{The \lhcb simulation application, Gauss: Design, evolution and experience}}, }{}\href{https://doi.org/10.1088/1742-6596/331/3/032023}{J.\ Phys.\ Conf.\ Ser.\  \textbf{331} (2011) 032023}\relax
\mciteBstWouldAddEndPuncttrue
\mciteSetBstMidEndSepPunct{\mcitedefaultmidpunct}
{\mcitedefaultendpunct}{\mcitedefaultseppunct}\relax
\EndOfBibitem
\bibitem{Blatt:1952ije}
J.~M. Blatt and V.~F. Weisskopf, {\em {Theoretical nuclear physics}}, \href{https://doi.org/10.1007/978-1-4612-9959-2}{ Springer, New York, 1952}\relax
\mciteBstWouldAddEndPuncttrue
\mciteSetBstMidEndSepPunct{\mcitedefaultmidpunct}
{\mcitedefaultendpunct}{\mcitedefaultseppunct}\relax
\EndOfBibitem
\bibitem{Yang:1950rg}
C.-N. Yang, \ifthenelse{\boolean{articletitles}}{\emph{{Selection rules for the dematerialization of a particle into two photons}}, }{}\href{https://doi.org/10.1103/PhysRev.77.242}{Phys.\ Rev.\  \textbf{77} (1950) 242}\relax
\mciteBstWouldAddEndPuncttrue
\mciteSetBstMidEndSepPunct{\mcitedefaultmidpunct}
{\mcitedefaultendpunct}{\mcitedefaultseppunct}\relax
\EndOfBibitem
\bibitem{Wilks:1938dza}
S.~S. Wilks, \ifthenelse{\boolean{articletitles}}{\emph{{The large-sample distribution of the likelihood ratio for testing composite hypotheses}}, }{}\href{https://doi.org/10.1214/aoms/1177732360}{Ann.\ Math.\ Stat.\  \textbf{9} (1938) 60}\relax
\mciteBstWouldAddEndPuncttrue
\mciteSetBstMidEndSepPunct{\mcitedefaultmidpunct}
{\mcitedefaultendpunct}{\mcitedefaultseppunct}\relax
\EndOfBibitem
\bibitem{LHCb-DP-2013-002}
LHCb collaboration, R.~Aaij {\em et~al.}, \ifthenelse{\boolean{articletitles}}{\emph{{Measurement of the track reconstruction efficiency at LHCb}}, }{}\href{https://doi.org/10.1088/1748-0221/10/02/P02007}{JINST \textbf{10} (2015) P02007}, \href{http://arxiv.org/abs/1408.1251}{{\normalfont\ttfamily arXiv:1408.1251}}\relax
\mciteBstWouldAddEndPuncttrue
\mciteSetBstMidEndSepPunct{\mcitedefaultmidpunct}
{\mcitedefaultendpunct}{\mcitedefaultseppunct}\relax
\EndOfBibitem
\bibitem{LHCb-PUB-2016-021}
L.~Anderlini {\em et~al.}, \ifthenelse{\boolean{articletitles}}{\emph{{The PIDCalib package}}, }{} \href{http://cdsweb.cern.ch/search?p=LHCb-PUB-2016-021&f=reportnumber&action_search=Search&c=LHCb+Notes} {LHCb-PUB-2016-021}, 2016\relax
\mciteBstWouldAddEndPuncttrue
\mciteSetBstMidEndSepPunct{\mcitedefaultmidpunct}
{\mcitedefaultendpunct}{\mcitedefaultseppunct}\relax
\EndOfBibitem
\bibitem{ATLAS:2014ala}
ATLAS collaboration, G.~Aad {\em et~al.}, \ifthenelse{\boolean{articletitles}}{\emph{{Measurement of $\chi_{c1}$ and $\chi_{c2}$ production with $\sqs = 7$\tev $pp$ collisions at ATLAS}}, }{}\href{https://doi.org/10.1007/JHEP07(2014)154}{JHEP \textbf{07} (2014) 154}, \href{http://arxiv.org/abs/1404.7035}{{\normalfont\ttfamily arXiv:1404.7035}}\relax
\mciteBstWouldAddEndPuncttrue
\mciteSetBstMidEndSepPunct{\mcitedefaultmidpunct}
{\mcitedefaultendpunct}{\mcitedefaultseppunct}\relax
\EndOfBibitem
\bibitem{LHCb-DP-2023-003}
LHCb collaboration, R.~Aaij {\em et~al.}, \ifthenelse{\boolean{articletitles}}{\emph{{Momentum scale calibration of the LHCb spectrometer}}, }{}\href{https://doi.org/10.1088/1748-0221/19/02/P02008}{JINST \textbf{19} (2024) P02008}, \href{http://arxiv.org/abs/2312.01772}{{\normalfont\ttfamily arXiv:2312.01772}}\relax
\mciteBstWouldAddEndPuncttrue
\mciteSetBstMidEndSepPunct{\mcitedefaultmidpunct}
{\mcitedefaultendpunct}{\mcitedefaultseppunct}\relax
\EndOfBibitem
\end{mcitethebibliography}

\clearpage
\centerline
{\large\bf LHCb collaboration}
\begin
{flushleft}
\small
R.~Aaij$^{38}$\lhcborcid{0000-0003-0533-1952},
A.S.W.~Abdelmotteleb$^{58}$\lhcborcid{0000-0001-7905-0542},
C.~Abellan~Beteta$^{52}$\lhcborcid{0009-0009-0869-6798},
F.~Abudin\'en$^{60}$\lhcborcid{0000-0002-6737-3528},
T.~Ackernley$^{62}$\lhcborcid{0000-0002-5951-3498},
A. A. ~Adefisoye$^{70}$\lhcborcid{0000-0003-2448-1550},
B.~Adeva$^{48}$\lhcborcid{0000-0001-9756-3712},
M.~Adinolfi$^{56}$\lhcborcid{0000-0002-1326-1264},
P.~Adlarson$^{86}$\lhcborcid{0000-0001-6280-3851},
C.~Agapopoulou$^{14}$\lhcborcid{0000-0002-2368-0147}
% ...

{\footnotesize \it

% ...
$^{14}$Universit{\'e} Paris-Saclay, CNRS/IN2P3, IJCLab, Orsay, France\\
% ...
$^{38}$Nikhef National Institute for Subatomic Physics, Amsterdam, Netherlands\\
% ...
$^{48}$Instituto Galego de F{\'\i}sica de Altas Enerx{\'\i}as (IGFAE), Universidade de Santiago de Compostela, Santiago de Compostela, Spain\\
% ...
$^{52}$Physik-Institut, Universit{\"a}t Z{\"u}rich, Z{\"u}rich, Switzerland\\
% ...
$^{56}$H.H. Wills Physics Laboratory, University of Bristol, Bristol, United Kingdom\\
% ...
$^{58}$Department of Physics, University of Warwick, Coventry, United Kingdom\\
% ...
$^{60}$School of Physics and Astronomy, University of Edinburgh, Edinburgh, United Kingdom\\
$^{61}$School of Physics and Astronomy, University of Glasgow, Glasgow, United Kingdom\\
$^{62}$Oliver Lodge Laboratory, University of Liverpool, Liverpool, United Kingdom\\
% ...
$^{70}$Syracuse University, Syracuse, NY, United States\\
% ...
$^{86}$Department of Physics and Astronomy, Uppsala University, Uppsala, Sweden, associated to $^{61}$\\
% ...
}
\end{flushleft}

\end{document}